\documentclass[fleqn,usenatbib]{mnras}

\usepackage{newtxtext,newtxmath}
\usepackage[T1]{fontenc}

\usepackage{threeparttable}
\DeclareRobustCommand{\VAN}[3]{#2}
\let\VANthebibliography\thebibliography
\def\thebibliography{\DeclareRobustCommand{\VAN}[3]{##3}\VANthebibliography}

\usepackage{graphicx}	% Including figure files
\usepackage{amsmath}	% Advanced maths commands
\title[Cloud-cloud collision in S235]{Cloud-cloud collision in S235: triggered the formation of high-mass stars and young star clusters}

\author[E. Chen et al.]{
En Chen,$^{1,2}$\thanks{E-mail: chenen@pmo.ac.cn}
Yu Gao,$^{1,3}$
Shiyu Zhang,$^{1}$
Xuepeng Chen,$^{1}$
Min Fang,$^{1}$
Qianru He$^{1,2}$
Xuejian Jiang,$^{4}$
\newauthor{
Yan Sun,$^{1}$
Xiaolong Wang,$^{1}$
and Hongjun Ma$^{1}$
}
\\
$^{1}$Purple Mountain Observatory $\&$ Key Laboratory of Radio Astronomy, No.10 Yuanhua Road, Nanjing, 210034, People's Republic of China\\
$^{2}$University of Science and technology of China, No.96 Jinzhai Road, Hefei, 230026, People's Republic of China\\
$^{3}$Department of Astronomy, Xiamen University, No.422 Siming South Road, Xiamen, Fujian, 361005, People's Republic of China \\
$^{4}$Research Center for Intelligent Computing Platforms, Zhejiang Laboratory, Hangzhou, 311100, People's Republic of China
}

\date{Accepted 2023 November 6. Received 2023 November 2; in original form 2023 May 19.}

\pubyear{2015}

\begin{document}
\label{firstpage}
\pagerange{\pageref{firstpage}--\pageref{lastpage}}
\maketitle

\begin{abstract}
We present the analysis of cloud-cloud collision (CCC) process in the Galactic molecular complex S235. 
Our new CO observations performed with the PMO-13.7m telescope reveal two molecular clouds, namely the S235-Main and the S235-ABC, with $\sim$ 4 km s$^{-1}$ velocity separation. 
The bridge feature, the possible colliding interface and the complementary distribution of the two clouds are significant observational signatures of cloud-cloud collision in S235. 
The most direct evidence of cloud-cloud collision process in S235 is that the S235-Main (in a distance of 1547$^{+44}_{-43}$ pc) and S235-ABC (1567$^{+33}_{-39}$ pc) meet at almost the same position (within 1$\sigma$ error range) at a supersonic relative speed. 
We identified ten $^{13}$CO clumps from PMO-13.7m observations, 22 dust cores from the archival SCUBA-2 data, and 550 YSOs from NIR-MIR data. 
63\% of total YSOs are clustering in seven MST groups (M1$-$M7). 
The tight association between the YSO groups (M1 $\&$ M7) and the bridge feature suggests that the CCC process triggers star formation there. 
The collisional impact subregion (the South) shows $3\sim5$ times higher CFE and SFE (average value of 12.3\% and 10.6\%, respectively) than the non-collisional impact subregion (2.4\% and 2.6\%, respectively), suggesting that the CCC process may have enhanced the CFE and SFE of the clouds compared to those without collision influence. 
%We also inferred that the S235-Main and the S235-ABC began to collide with each other in $3\sim5$ Myr ago and then produced large amounts of compressed gas to form high-mass stars in $\sim$0.5 Myr ago. Until today, the two clouds may be pulling out of collision and moving away from each other.
\end{abstract}


%%%%%%%%%%%%%%%%%%%%%%%%%%%%%%%%%%%%%%%%%%%%%%%%%%

%%%%%%%%%%%%%%%%% BODY OF PAPER %%%%%%%%%%%%%%%%%%

\section{Introduction}

Star formation activity is one of the most fundamental processes controlling galactic evolution. 
%Stars are believed to be born from dense gas with an tight correlation (a slope of $\sim$1) found by \cite{Gao2004}. 
Stars are believed to be born from dense molecular gas. A tight correlation between the mass of dense molecular gas and the star formation rate (SFR) was found by \cite{Gao2004}, known as the law of star formation (or the Kennicutt-Schmidt relation \citep{Kennicutt1998}).
In recent theories, the star formation process is thought to be due to gravitational instability in a turbulent molecular cloud that causes the cloud to collapse and thus form stars \citep{Mckee2007, Zinnecker2007}. High density is one of the essential conditions for molecular clouds to collapse. There are many mechanisms that can compress gas to achieve high density condition, such as transporting material through filament accretion \citep{Peretto2013, Motte2015, Motte2017}, compressing gas through stellar feedback (such as supernova feedback \citep{Herbst1978}, massive star ionized hydrogen feedback \citep{Whitworth1994}, etc.) and dynamic feedback (such as cloud-cloud collision process \citep{Habe1992}). All of these mechanisms can explain the formation of dense regions in galaxies, which are hotbeds for star formation. 

Massive star formation has attracted much attention in the past few decades because it plays an important role in galaxy formation and evolution \citep{Mckee2003}. However, the mechanism of which is a long-standing problem. The difficulty of massive star formation lies in achieving their initial conditions (ultra-high density) and balancing their strong feedback (radiation pressure greater than gravitational potential). The cloud-cloud collision (CCC) process works well to address the ultra-high density conditions \citep{Takahira2018}. Collisions between two or more molecular clouds can produce supersonic shock waves in collisional area, which can greatly compress gas, allowing large amounts of matter to be concentrated in small-scale structures for a short period of time (typical compression timescale of 0.1 $\sim$ 0.5 Myr), and eventually collapse to form massive stars. Star formation triggered by dynamic feedback can effectively regulate the star formation rate of galaxies, such as the formation of super star clusters (SSC) or mini-starbursts in the local region of galaxies \citep{Fukui2021}, where the contribution of cloud-cloud collision cannot be ignored. Cloud-cloud collision process occur very frequently in galaxies \citep{Tasker2009}, and 10\% (up to half) of the galaxy's total star formation are caused by such process \citep{Kobayashi2018}.

A pioneering simulation work of CCC was performed by \cite{Habe1992}, in which a basic scenario of CCC was presented. 
%They indicated  that a dense gas layer can be formed by the compression of the supersonic collision where it induce the formation of dense self-gravitating clumps which may finally evolve into high-mass stars.
They advanced that a dense gas layer can be facilitated by the compression from supersonic collision, and it induce the formation of dense self-gravitating clumps which may finally evolve into clusters containing high-mass stars.
CCC is an effective trigger of massive, dense cores formation. In the recent hydrodynamical simulations \citep[e.g.,][]{Takahira2014, Takahira2018}, a bulk of cores are born in the shocked regions and the number and mass of cores are controlled by the relative collision speeds (typically at 10 km s$^{-1}$). 
%A faster relative velocity increases \textbf{(could increase/ suppress)} the number of cores but suppresses the mass growth of cores due to the short collision time for cores to accrete gas.
A larger relative speed could increase the number of cores, but on the contrary, it suppresses the mass growth of cores due to short collision timescale for cores to accrete gas. 
%Therefore, collisions with different initial conditions (such as size, collision speeds) of two colliding clouds may result in a wide range of stellar mass of stars. 
Therefore, collisions of two clouds with different initial conditions (such as size, mass and collision speeds) may result in a wide range of stellar mass, for example, a single or several O stars triggered by smaller molecular clouds (M $> 10^{3-5}$ M$_{\odot}$) collisions, including L1188 \citep{Gong2017a}, N159W-S in LMC \citep{Fukui2015}, Trifid Nebula M20 \citep{Torii2011, Torii2017}, RCW 120 \citep{Torii2015}, S235 \citep{Dewangan2017}, Sh2-48 \citep{Torii2017a}, and RCW 34 \citep{Hayashi2018}; a Super Star Cluster (SSC) or a mini-starburst triggered by massive molecular cloud (M $> 10^{5-6}$ M$_{\odot}$), including Westerlund 2 \citep{Furukawa2009}, NGC3603 \citep{Fukui2014}, RCW38 \citep{Fukui2016}, W51A \citep{Fujita2017}, and NGC6334/NGC6357 \citep{Fukui2018a}.

Among these observations, the 'bridge feature' shown in position-velocity diagrams was believed to be the commonest signature of CCC. 
The bridge feature is the coherent gas which connects the two colliding clouds and the turbulent motion inside would be enhanced by collisions %\textbf{(better to divide into two sentences because there is no logical relationship before and after and)}
\citep[][]{Wu2017a}. This feature can also be reproduced by the numerical simulations calculated by \cite{Anathpindika2010} and \cite{Takahira2014}. Other observational signatures such as the complementary spatial distribution of adjacent clouds and the cavity created on the larger cloud are also very important evidences to prove a collision event. However these signatures could not always be seen in a particular collision event due to the variant collision stages and different projection along the line-of-sight. \cite{Fukui2018a} summarized previous studies  and listed the main observational signatures of CCC as: a) the supersonic velocity separation of the colliding clouds, b) the complementary distribution of them, and c) the bridge feature connecting clouds. These signatures are useful guidance to search more collision candidates in spectroscopic observations. 

\begin{figure}
\includegraphics[width=\columnwidth]{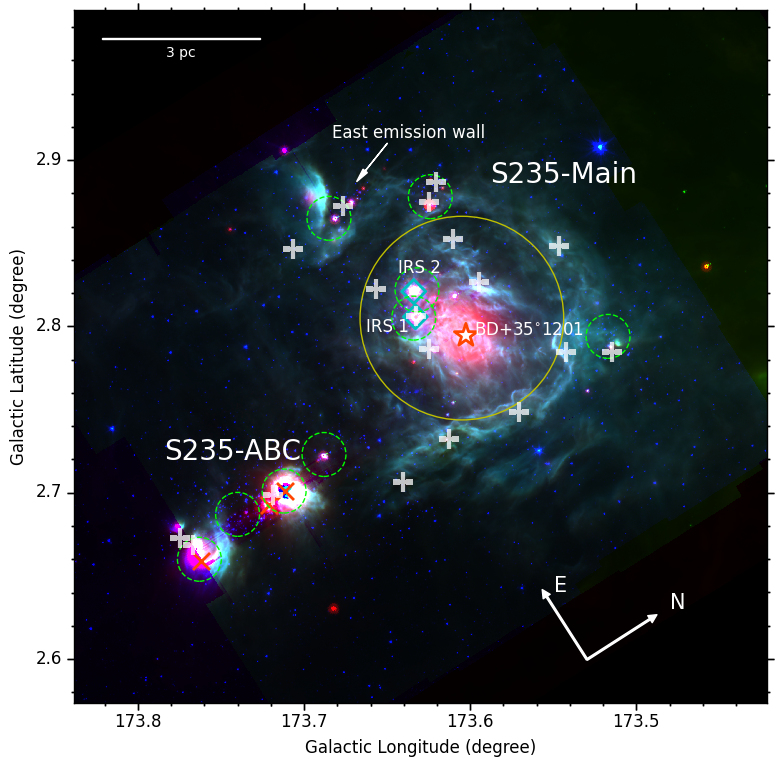}
\caption{Color composite image of the S235 complex. Blue, green and red show 
the $Spitzer$ IRAC 3.6 $\mu$m, 8 $\mu$m and $Spitzer$ MIPS 24 $\mu$m, respectively. 
The complex mainly hosts two molecular clouds, the S235-Main and the S235-ABC. The red asterisk symbol indicates the exciting O star BD+35$^{\circ}$1201 of the S235 \textsc{Hii} region (the yellow ring) and the red crosses indicate B type stars in S235-ABC. The cyan diamonds indicate the two bright infrared sources IRS 1 and IRS 2. The green dashed circles indicate the 
position of nine young star clusters. The white pluses indicate 
the Bolocam clumps at 1.1 mm. A nebular emission is designated as eastern emission wall which is far away from the exciting O star. \label{fig:rgb}
}
\end{figure}

The S235 complex, known as an active star-forming complex, is part of the giant molecular cloud G174+2.5 in the Perseus Spiral Arm \citep{Heyer1996} toward the anti-Galactic Plane ($l,b$)$\sim$(173$^{\circ}$.6, 2$^{\circ}$.8). Figure \ref{fig:rgb} shows the color-composite images of the S235 region that was obtained by $Spitzer$ IRAC 3.6 $\mu$m (blue) and 8 $\mu$m (green), and $Spitzer$ MIPS 24 $\mu$m (red) images. The molecular complex mainly hosts the S235 \textsc{Hii} region (the yellow ring) with a diameter of $\sim$ 3 pc and a slight faint cloud with three ultra-compact (UC) \textsc{Hii} regions (S235-A, S235-B and S235-C) on it in the south . The S235 \textsc{Hii} region who was excited by a single O9.5V type star BD+35$^{\circ}$1201 \citep{Georgelin1973}, was surrounded by a dust ring with an opening end toward the south. The distance of S235 was estimated to be 1.8 kpc by \cite{Evans1981}.
S235 is a very well studied site using multi-wavelength datasets spanning from near-infrared (NIR) to radio wavelengths. Over a hundred of young stellar objects (YSOs) were found in the extended S235 region \citep[e.g.,][]{Dewangan2011, Dewangan2016} and totally nine young embedded clusters (ECs) were identified in some well known star forming sites called East 1, East 2, North-West, Central and S235A, S235B and S235C, suggesting a sequential star formation in the molecular complex \citep{Camargo2011}. 
 %\textbf{(Why not putting all the HII region introduction at above text??)}

In the early studies, \cite{Evans1981} found two main velocity components ($-$20 km $s^{-1}$ and $-$17 km $s^{-1}$) toward the  S235 region. We hereafter refer to the $-$20 km $s^{-1}$ cloud as the "S235-Main" and the $-$17 km $s^{-1}$ cloud as the "S235-ABC". The S235-Main cloud was associated with an evolved \textsc{Hii} region (S235 \textsc{Hii} region) with a nearly sphere-like shell (a diameter of $\sim$ 3 pc), harboring at least four star forming sites (i.e., East 1, East 2, North-West, and Central). Furthermore, the S235-Main was constructed by three gas components in details (i.e., $-$18 $\sim$ $-$15 km $s^{-1}$ (red), $-$21 $\sim$ $-18$ km $s^{-1}$ (central), and $-25$ $\sim$ $-21$ km $s^{-1}$ (blue)) traced by $^{13}$CO ($1-0$) line data \citep{Kirsanova2008}, which shows a C-like structure in position-velocity diagram, indicating an expanding motion of the S235 \textsc{Hii} region \citep{Dewangan2016}. The spatial distribution of the star forming sites and the kinematics of molecular gas in the S235-Main cloud suggest that the expansion of the S235 \textsc{Hii} region could be responsible for the star formation around the shell. On the other hand, the S235-ABC cloud only harbors three star forming sites (i.e., S235 A, S235 B and S235 C), embedding with three B type stars respectively. The sites of these stars are located at about 5$-$6 pc south of the S235-Main which is larger than the diameter of the S235 \textsc{Hii} region, suggesting that their star formation could not be triggered by the expansion of the S235 \textsc{Hii} region ionization front \citep{Camargo2011}. However, several observational signatures such as broad bridge feature in velocity phase and complementary molecular pair of S235-Main and S235-ABC were found by \cite{Dewangan2017a} using FCRAO $^{12}$CO ($1-0$) and $^{13}$CO ($1-0$) datasets suggesting that the two clouds were suffering the CCC process. Their analysis revealed that the CCC process might have influenced the star formation activity along the colliding interface (i.e., East 1 and S235-ABC), which triggered the formation of massive stars and young stellar clusters in S235. 

Although the CCC signatures are significant toward the S235 region, the connection between the CCC process and the star formation activity, especially its effects on triggering star formation in the two colliding clouds has not been studied. In this paper we present the results of our new observations of $^{12}$CO ($1-0$), $^{13}$CO ($1-0$) and C$^{18}$O ($1-0$) with angular resolutions of $52^{\prime\prime}$ (at 115 GHz) to $55^{\prime\prime}$ (at 110 GHz) using the PMO-13.7m telescope. The data was also combined with other archived data such as parallax observations from the Gaia satellite \citep{Gaia2016}, NIR data from the UKIDSS Galactic Plane Survey (UKIDSS-GPS; \citealt{Lawrence2007}), NIR and MIR data from $Spitzer$, dust continuum 1.1 mm data from the Bolocam Galactic Plane Survey (BGPS; \citealt{Aguirre2011}), and radio continuum data from the Canadian Galactic Plane Survey (CGPS; \citealt{English1998}) to investigate the detailed scenario of CCC in S235 and its effects on triggering star/core formation. Section \ref{sec:obser} describes the observations and data reductions and Section \ref{sec:results} presents the observation results. Discussions are given in Section \ref{sec:discussion} and conclusions are placed in Section \ref{sec:conclusions}.

\section{Observations} \label{sec:obser}

\subsection{PMO-13.7m CO observations\label{subsec:pmo}}

We employed the PMO-13.7m single-dish sub-millimeter telescope to study the distributions and kinematics of the molecular material associated with the S235 complex. The observations of S235 were made as part of the Milky Way Imaging Scroll Painting (MWISP\footnote{\url{http://www.radioast.nsdc.cn/mwisp.php/}}) project which is aimed at mapping the large-scale molecular gas with CO and its isotope molecular lines along the northern Galactic Plane. The project is conducted by the PMO-13.7m telescope, which is located at Delingha (3200m altitude), Qinghai, China. This antenna can observe three CO molecular lines, $^{12}$CO ($1-0$), $^{13}$CO ($1-0$), and C$^{18}$O ($1-0$) simultaneously. The 3$\times$3-beam Superconducting Spectroscopic Array Receiver (SSAR) system is used as front end, which provides a 1 GHz bandwidth with 16384 channels and a spectral resolution of 61 kHz, equivalent to a velocity coverage of $\sim$ 2600 km s$^{-1}$ and a velocity resolution of $\sim$ 0.17 km s$^{-1}$ at 110 GHz. The detailed properties of this system are described in \cite{Shan2012}. In order to observe the three low-J CO lines simultaneously, the $^{12}$CO ($1-0$) line was set at the center of the upper sideband so that the lower sideband is able to cover both $^{13}$CO ($1-0$) and C$^{18}$O ($1-0$) lines. Observations were undertaken in position-switch on-the-fly (OTF) mapping mode. The observed region was divided into several $30^{\prime}\times30^{\prime}$ cells and each cell was scanned at least in two orthogonal directions, along the galactic longitude and the galactic latitude, to reduce scanning effects. 

According to the status report\footnote{\url{http://english.dlh.pmo.cas.cn/fs/}} of the PMO-13.7m telescope, the beam efficiency $B_{eff}$, which is used to convert the antenna temperature ($T_A$) to the main beam temperature with the relation $T_{mb}=T_A/B_{eff}$, can reach to 46\% at 115 GHz and 49\% at 110GHz. The receiver equipped with on the telescope provides typical system temperatures of $191-381$ K at the upper sideband, and $142-237$ K at the lower sideband. Therefore, the typical rms noise level is about 0.5 K ($T_{mb}$) for $^{12}$CO ($1-0$), and 0.3 K ($T_{mb}$) for $^{13}$CO ($1-0$) and C$^{18}$O ($1-0$) at a channel width of $\sim$0.17 km s$^{-1}$. The telescope has a beam size of 55$^{\prime\prime}$ at 110 GHz and 52$^{\prime\prime}$ at 115 GHz, and the pointing of the telescope has an rms accuracy of about 5$^{\prime\prime}$.

Data reduction was done with CLASS and GreG in the software GILDAS\footnote{\url{https://www.iram.fr/IRAMFR/GILDAS/}}. The final data product was convert into fits file format with pixel size of 30$^{\prime\prime}$ $\times$ 30$^{\prime\prime}$. %\sout{Finally, we clipped the study region of S235 from the data cube covered} 
A field of 25$^{\prime} \times$25$^{\prime}$ centred at ($l, b$) $\sim$ (173$^{\circ}$.625, 2$^{\circ}$ .790) is taken in this work. Throughout this paper, the galactic coordinate system is utilized and the equinox is J2000.0, and velocities are all given with respect to the local standard of rest (LSR). 

%\subsection{JCMT high-J CO observations \label{subsec:jcmt}}

\subsection{Archival data\label{subsec:arx}}

We obtained the near-infrared (NIR) J, H, K band point sources catalog from the UKIDSS-GPS data release \citep{Lawrence2007} and the selection procedures of the UKIDSS-GPS photometry were following \cite{Dewangan2015}. Only brighter NIR sources (H$<12.3$ and K$<11.4$  mag) were extracted from 2MASS \citep{Skrutskie2006} to take the place of the sources from UKIDSS-GPS in order to avoid saturation. Using the NIR data in combination with the $Spitzer$ MIR data \citep{Fazio2004}, we searched for the disk-bearing young stellar object (YSO) candidates in the surveyed region (details are described in Section \ref{ysos}). We made use of the Bolocam Galactic Plane Survey (BGPS) of 1.1 mm dust continuum emission \citep{Rosolowsky2010, Aguirre2011, Ginsburg2013} and the available nightly observations of James Clerk Maxwell Telescope (JCMT) Submillimeter Common-User Bolometer Array 2 (SCUBA-2; \citealt{Dempsey2013, Chapin2013}) of 850 $\mu$m emission (Project ID: MJLSY02) to study the dust clumps and cores (more details are described in Section \ref{subsec:clump}). The angular resolution of the 1.1 mm map is $\sim$33$^{\prime\prime}$ while the 850 $\mu$m map has a higher resolution of $\sim$14$^{\prime\prime}$. The 1420 MHz radio continuum data were obtained from the Canadian Galactic Plane Survey (CGPS; \citealt{English1998}) with beam size of $\sim$ 1$^{\prime}$. In order to calibrate the distance of the S235 molecular cloud, we also utilized the parallax data from Gaia data release 3 \citep{Gaia2021}.

\section{Results and Discussions} \label{sec:results}

\subsection{Molecular gas distributions with PMO-13.7m} \label{subsec:distribution}

We obtained $^{12}$CO (J$=1-0$), $^{13}$CO (J$=1-0$) and C$^{18}$O (J$=1-0$) line emissions simultaneously by PMO-13.7m telescope. 
Two major molecular clouds, namely the S235-Main and the S235-ABC, were revealed base on the gas kinetics motion. 
Our CO data has a similar spatial resolution with the FCARO CO data. With the CO data, we can reproduce the results of cloud-cloud collision in S235 which has been presented in \cite{Dewangan2017a}. Our data has better velocity resolution (0.17 km s$^{-1}$) than FCRAO data (0.266 km s$^{-1}$), which can reveal more detailed velocity components, such as bridge features with coherent gas. Additional observation of C$^{18}$O (J$=1-0$) could reveal more clumpy regions associated with dense regions (or star formation regions).

\begin{figure*}
\includegraphics[width=\textwidth]{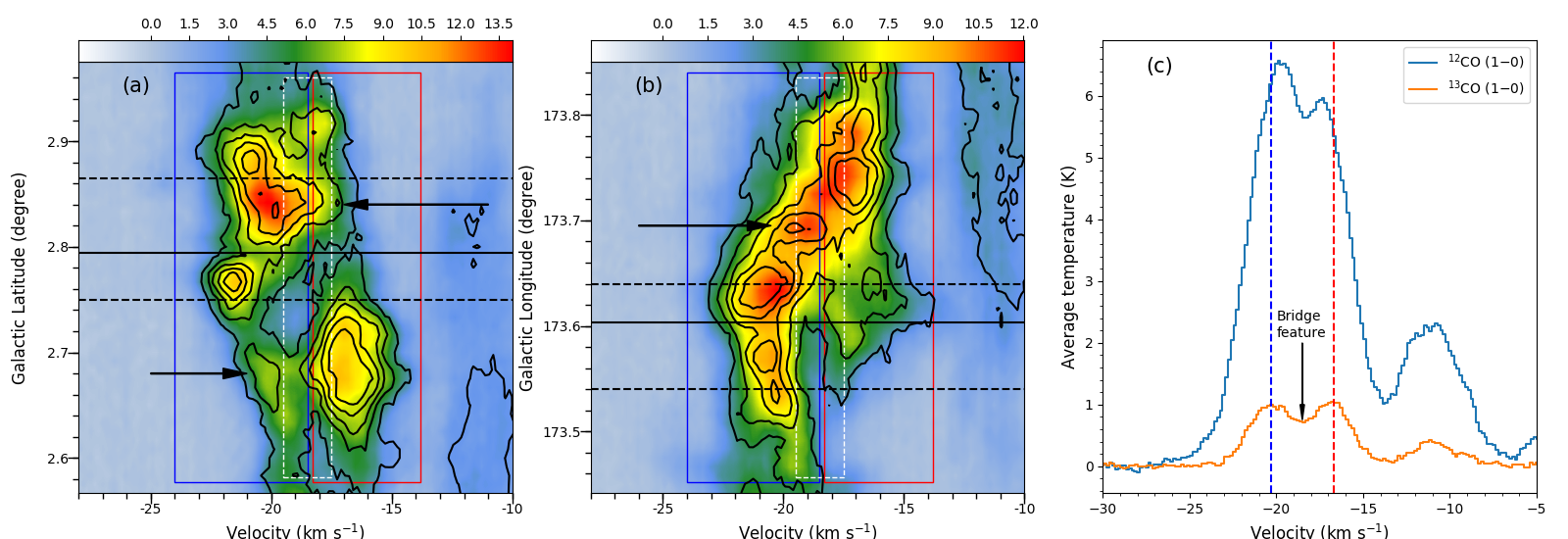}
\caption{Position-velocity diagrams of the S235 complex. Contours in (a) and (b) show the averaged intensity 
of the PMO-13.7m $^{13}$CO (1-0) data while the background color image shows that of the $^{12}$CO (1-0) data.The horizontal black solid line and dashed lines indicate the position of the exciting O star and the approximate extension of the 8 $\mu$m ring, respectively. The blue and red rectangle highlight the velocity range of the S235-Main cloud ($-24$ $\sim$ $-18.5$ km s$^{-1}$) and  the S235-ABC cloud 
($-18.5$ $\sim$ $-13.5$ km s$^{-1}$) , respectively, while the white dashed rectangle highlights the intermediate velocity range ($-19.5$ $\sim$ $-17.5$ km s$^{-1}$) of the bridge feature which was significantly influenced by collision. %, and it's spatial distribution was shown in Figure\ref{fig:yso_groups}. 
(c) shows the average spectrum of the whole region. The black arrows in all panels represent the bridge feature that connects the two separated clouds.   \label{fig:pvmap}}
\end{figure*}

\iffalse
As did in \cite{Dewangan2017a}, we present the the zeroth (moment 0), first (moment 1) and second (moment 2) moment maps in Figure \ref{fig:moments}a\&b\&c, respectively, tracing the signature of cloud-cloud collision in S235. These moment maps are integrated over the velocity window of (-24, -15) km s$^{-1}$, which reveal the integrated-intensity, velocity and velocity dispersion of the gas. The first moment map clearly trace two distinct molecular clouds (i.e. the S235-Main and the S235-ABC) with perfect complementary distribution in boundary. We highlight this boundary with purple dashed line in Figure \ref{fig:moments}b and put it to Figure \ref{fig:moments}a\&c. The boundary present large velocity dispersion ($\delta$V> 2 km s$^{-1}$) in Figure \ref{fig:moments}c, indicating that this boundary is a possible colliding interface of the two clouds. 
\fi

Figure \ref{fig:pvmap} show the position-velocity (PV) diagrams of $^{12}$CO (background) and $^{13}$CO (contours) toward the survey region. 
The S235-Main is the blue-shifted cloud with a velocity range from $-24$ to $-18.5$ km s$^{-1}$ and a mean velocity of $-20.5$ km s$^{-1}$, while the S235-ABC is the red-shifted cloud with a velocity range from $-18.5$ to $-13.5$ km s$^{-1}$ and a mean velocity of -16.5 km s$^{-1}$. Therefore, there is $\sim$ 4 km s$^{-1}$ velocity separation between the two clouds along light-of-sight. 
Note that the two clouds are not completely separated, they are connected to each other by bridge feature (black arrows in Figure \ref{fig:pvmap}), which has a velocity range of $-19.5$ to $-17.5$km s$^{-1}$. The bridge feature is the compressed gas layer decelerated by the collision, which transforms the kinetic motion of the molecular cloud into a turbulent motion with a broad velocity dispersion. The existence of a broad bridge feature in velocity space indicates an observational signature of collisions between molecular clouds \cite{Torii2017}.

On the other hand, we can also see a semi-ring-like or C-like structure in the blue-shifted velocity phase. 
As shown in Figure \ref{fig:pvmap}a\&b, the horizontal black solid line and the dashed lines indicate the position of the exciting O star of the S235 \textsc{Hii} region and the approximate extent of the 8 $\mu$m ring emission, respectively. 
This structure is a signature of an expanding shell of the \textsc{Hii} region, which is in agreement with the previous work of \cite{Dewangan2016}, which detected an expansion velocity of the gas to be $\sim$3 km s$^{-1}$. 
Therefore the S235-Main was mainly shaped by the S235 \textsc{Hii} region, while the S235-ABC may not be affected by the expansion shock of the S235 \textsc{Hii} region due to the larger separation velocity of 4 km s$^{-1}$.

\begin{figure*}
\includegraphics[width=\textwidth]{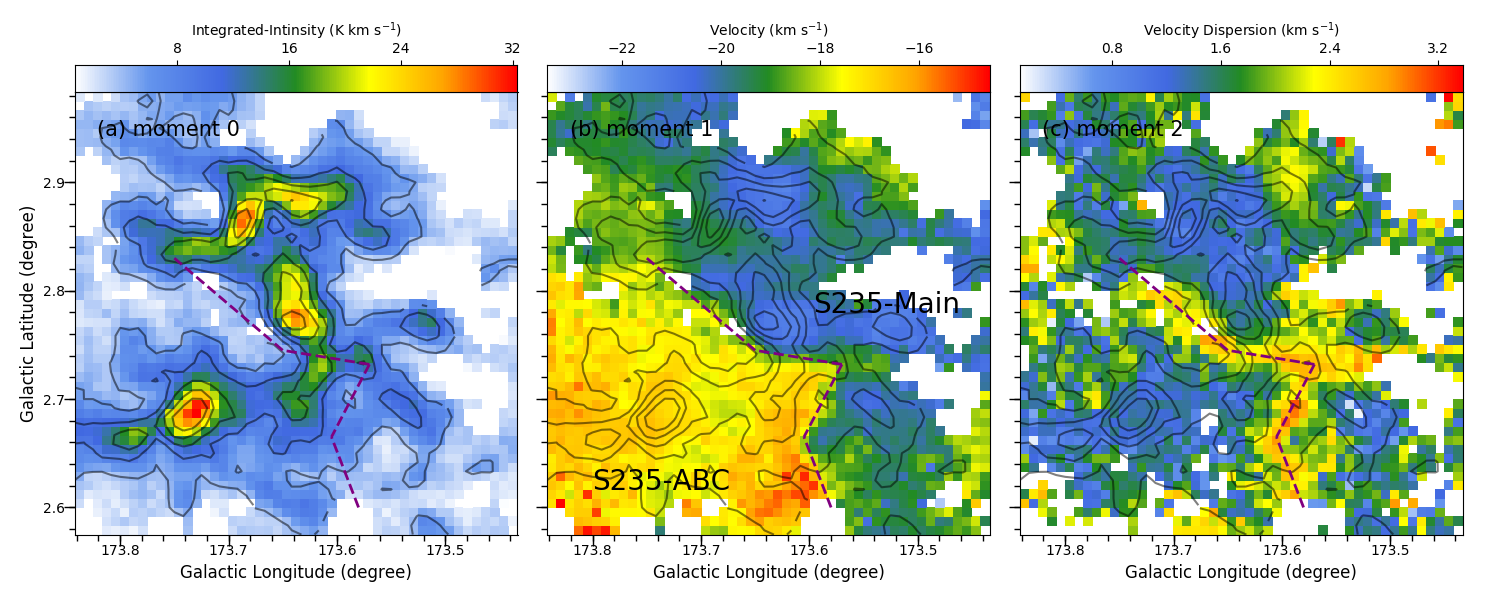}\\
%\vspace{-1.1cm}
\includegraphics[width=\textwidth]{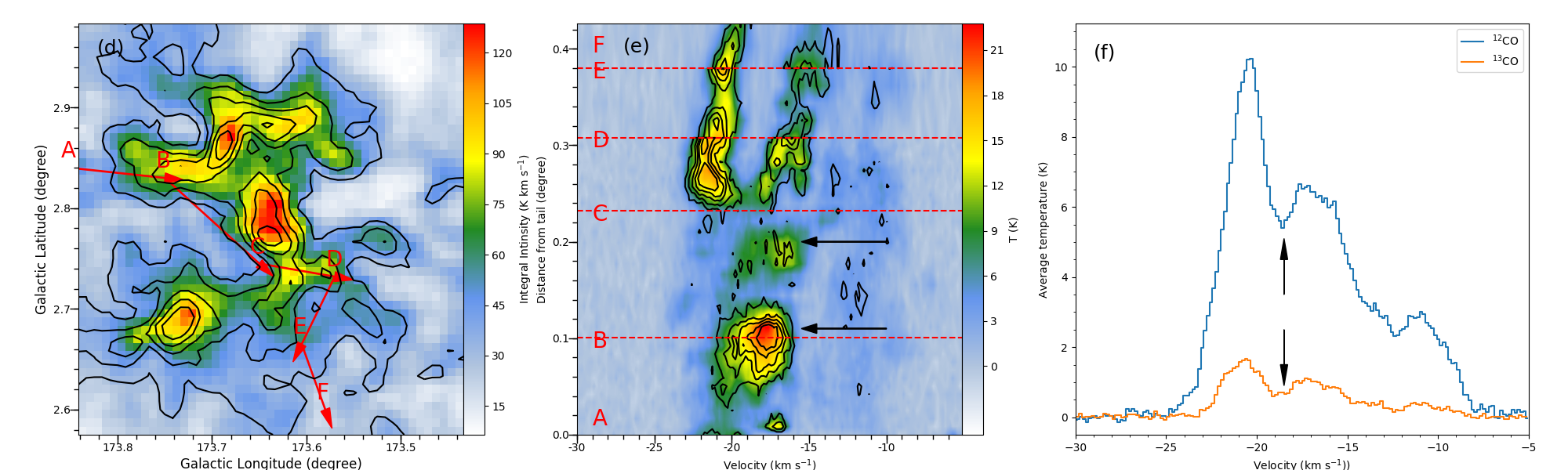}
\caption{(a)$-$(c) $^{13}$CO momentum maps of the S235 complex. The integrated velocity ranges from $-24$ 
km s$^{-1}$ to $-13.5$ km s$^{-1}$. Pixels where the maximum intensity value within that velocity range less than 0.9 K (3$\sigma$ noise) are omitted and set to blank. The moment 0, moment 1 and moment 2 maps are presented 
in the left, center and right respectively. The contours represent the integrated intensity of the $^{13}$CO whose intensity levels start from 5 K km s$^{-1}$(5.5 $\sigma$) to 30 K km s$^{-1}$ with steps of 5 K km s$^{-1}$. The purple dashed line highlights the possible collision 
interface between S235-Main and S235-ABC. (d) Five position-velocity cuts (PV-cuts) along the interface with markers A-F. (e) Position-velocity diagram of the PV-cuts. (f) Average spectrum along the interface. The black arrows in all panels depicted the bridge feature.  \label{fig:moments}
}
\end{figure*}

The observational signatures of cloud-cloud collision is S235 has been discussed in the work of \cite{Dewangan2017a}. They used the first and the second momentum maps of $^{12}$CO and $^{13}$CO to trace the boundary and colliding interface of the two clouds. 
As shown in Figure \ref{fig:moments}a$-$c, we reproduced the momentum maps for the $^{13}$CO (J=1-0) emission using our PMO-13.7m data. 
A distinct boundary between the S235-Main and the S235-ABC is found in the first momentum map (moment 1). The convex distribution in the north of S235-ABC just matches the concave distribution in the south of S235-Main. 
This perfect complementary distribution (concave-convex pairs) indicates that the S235-Main may have physical connection with the S235-ABC. 
Moreover, as shown in the second momentum map (moment 2), the boundary region has a velocity dispersion of $\sim$2 km s$^{-1}$ which nearly double the velocity dispersion of each cloud. 
This large velocity dispersion can be interpreted as the overlap or interaction of the two clouds in the line-of-sight. 
If two clouds are at the same distance, then there is an interaction between them, such as cloud-cloud collision. We will measure the distance of each cloud in Section \ref{sec:distance} to support this point. 
Therefore, a possible colliding interface (highlighted by the purple dashed line) can be found in the boundary of the two clouds. 
We also checked the position-velocity (PV) distribution along the colliding interface by five PV-cuts (A to F) in Figure \ref{fig:moments}d$-$f. 
PV-cuts from A to C show a broad single velocity component with an average velocity of $\sim-18$ km s$^{-1}$, which represents the intermediate velocity bridge feature that connects the S235-Main and the S235-ABC, indicating that the two clouds may have been interacted with each other and converted their dynamical motion into the turbulent motion. 
However, PV-cuts from C to F clearly show two velocity components that refer to the S235-Main and the S235-ABC velocity domain respectively, indicating that the two clouds may overlay in these regions in the line-of-sight or may be colliding or separating with each other in the near future. 
The averaged CO spectrum of the interface is show in Figure \ref{fig:moments}f, which present a bridge feature of flattened profile in the intermediate velocity range.

In conclusion, the supersonic velocity separation ($\sim$4 km s$^{-1}$), the spatial complementary distribution of the molecular cloud pairs and the broad velocity bridge feature along the colliding interface are distinct signatures of cloud-cloud collision process between the S235-Main and the S235-ABC, which is consistent with the results of \cite{Dewangan2017a}. Detail distance measurement of the two clouds will support this point.

\subsection{Molecular cloud identification and distance measurement}\label{sec:distance}

\begin{figure*}
    \centering
    \includegraphics[width=\textwidth]{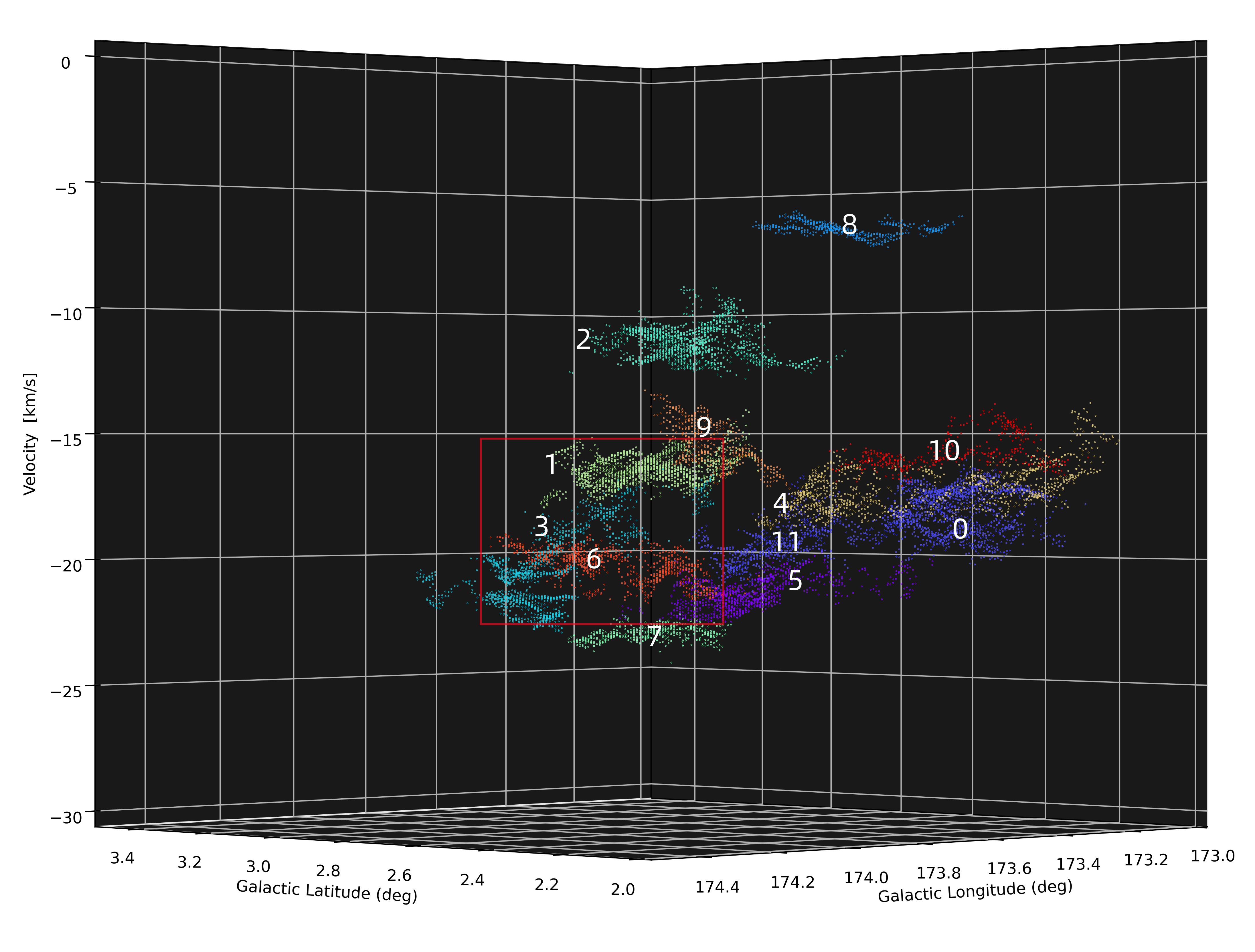}
    \caption{Three-dimensional presentation of molecular cloud clustering decomposition in the extended S235 region. The red box represents the extent of the S235 study area, which contains three components marked with label 1, 3 and 6.}
    \label{fig:clouds}
\end{figure*}

The distance of S235 is commonly used as 1.8 kpc in literature \citep[e.g.,][]{Evans1981}, but this value is not particularly strict. It was estimated based on the intermediate value of earlier studies such as the 2.1 kpc distance from the O-star determined by \cite{Georgelin1973} and the 1.6 kpc distance from the nebula estimated by \cite{Kazes1977}. 
We employed a method based on Gaia DR2 parallax and G-band extinction (A$_G$) measurement introduced by \cite{Yan2019b} to measure the distance of each velocity component in S235. This method made use of the boundary of molecular clouds and the estimation of parallax as well as the extinction of stars by Gaia satellite.

In the method, the first step is to identify the boundary of each molecular cloud by decomposing the $^{13}$CO lines into individual components with Gaussian fitting. Due to the spatial consistency of the molecular cloud, the pixel points with similar line-center velocity, similar spectral line broadening and intensity are clustered, so as to identify each cloud. In this work, we use ACORNS \footnote{\url{https://github.com/jdhenshaw/acorns.}} (Agglomerative Clustering for ORganising Nested Structures) clustering to distinguish clouds \citep{Henshaw2019}. ACORNS is a clustering tool based on a technique known as hierarchical agglomerative clustering, whose primary function is to generate a hierarchical clustering system in discrete data.

In the clustering results, each Tree corresponds to a cloud. For a Tree with hierarchical structure, the cloud is a complex composed of discrete sub-structures without overlapping in the projection phase. On the other hand, for a Tree without hierarchical structure, the cloud is a relatively uniform and coherent structure. The positions of the outermost pixels are connected to form the boundary of the cloud on the projection plane. In the line-of-sight direction, the Gaussian fitting line width of the corresponding position is taken as the thickness of the cloud. Consistent with this, the range of the line-center velocity within $\pm 3\sigma$ is taken as the velocity range of the cloud. Therefore, according to the clustering method mentioned above, the velocities of adjacent pixels are similar, which makes the cloud have a relatively smooth contour in the PPV space.

For the next step, we divide Gaia stars into two classes, i.e. on-cloud stars and off-cloud stars, according to the boundary of individual molecular cloud. Due to the high extinction across molecular clouds, the starlight will be attenuated when passing through the molecular cloud, so that the extinction value will significantly change between the foreground and background of the molecular cloud. Therefore, the parallax of the extinction jump point represents the true distance of the molecular cloud.

The final step is to build a Bayesian model to estimate the distance to the on-cloud stars, and solve the parameters in the model with Markov chain Monte Carlo (MCMC) sampling. The position of jump point (D), the extinction value of foreground and background (A$_{G}$) and its standard deviation ($\sigma$) were selected as priors, and the truncated Gaussian model was used as the probability density distribution. Considering that most of the distance of this area is within 3 kpc, and the extinction is very strong, we set the distance of the jump point to be evenly distributed from 100 to 3000 pc. The initial distance value is set to the average distance of the selected stars. Both the extinction value and its deviation are set as exponential distribution. The selection of initial value and the combination of errors follow the method from \cite{Yan2019a}.

Figure \ref{fig:clouds} shows the 3D PPV molecular cloud clustering of the extended S235 area ($173^{\circ}.0<l<174^{\circ}.5$, $2^{\circ}<b<3^{\circ}.5$). 
A total of 12 cloud components are decomposed in this region, among which the red box region is the main molecular cloud components in the S235 study area in the paper, including molecular clouds marked with label 1, 3 and 6 in the figure, which correspond to cloud-3, cloud-2 and cloud-1 in the latter part of the paper respectively. 
The properties of these three molecular components are listed in table \ref{tab:cloud_dist}.

\begin{figure*}
    \centering
    \includegraphics[width=\textwidth]{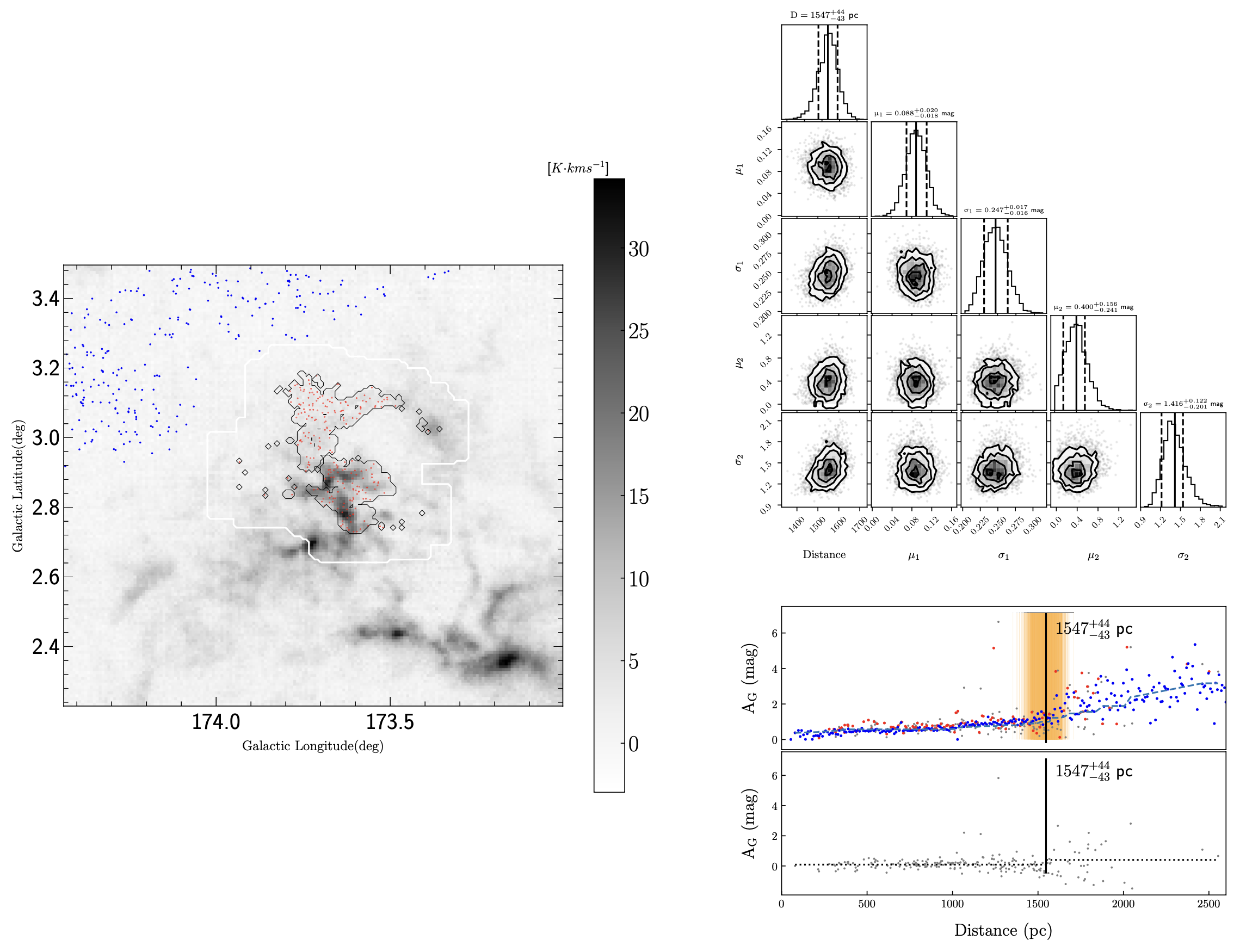}
    \caption{Distance measurement of the molecular cloud component cloud-1 in the S235 region. The gray scale on the left panel shows the $^{13}$CO integral intensity map for the selected data area. Black contour is the boundary of the velocity component $^{13}$CO molecular cloud found by cloud segmentation algorithm. White contour is the extension of the molecular cloud boundary contour. The red dots represent the position of on-cloud stars, while the blue dots represent the stars in the selected reference region, which exclude the region with obvious $^{12}$CO emission and the region within the white contour to reduce the effect of background extinction on the background subtraction. The upper right panel is the corner plot of MCMC sampling. The upper and lower subgraphs in the lower right panel respectively represent the $A_{G}$-$Distance$diagram before and after the baseline of background extinction fitting is subtracted. The data used for MCMC fitting is the data after baseline subtraction. The black vertical line represents the median distance obtained by sampling, the orange translucent thin line represents all the posterior sampling results, and the confidence interval with the upper and lower indices of the distance is 1$\sigma$. The error here is a mathematical statistical error and the actual error should be plus a systematic error of about 5\%.}
    \label{fig:cloud1}
\end{figure*}

\begin{figure*}
    \centering
    \includegraphics[width=\textwidth]{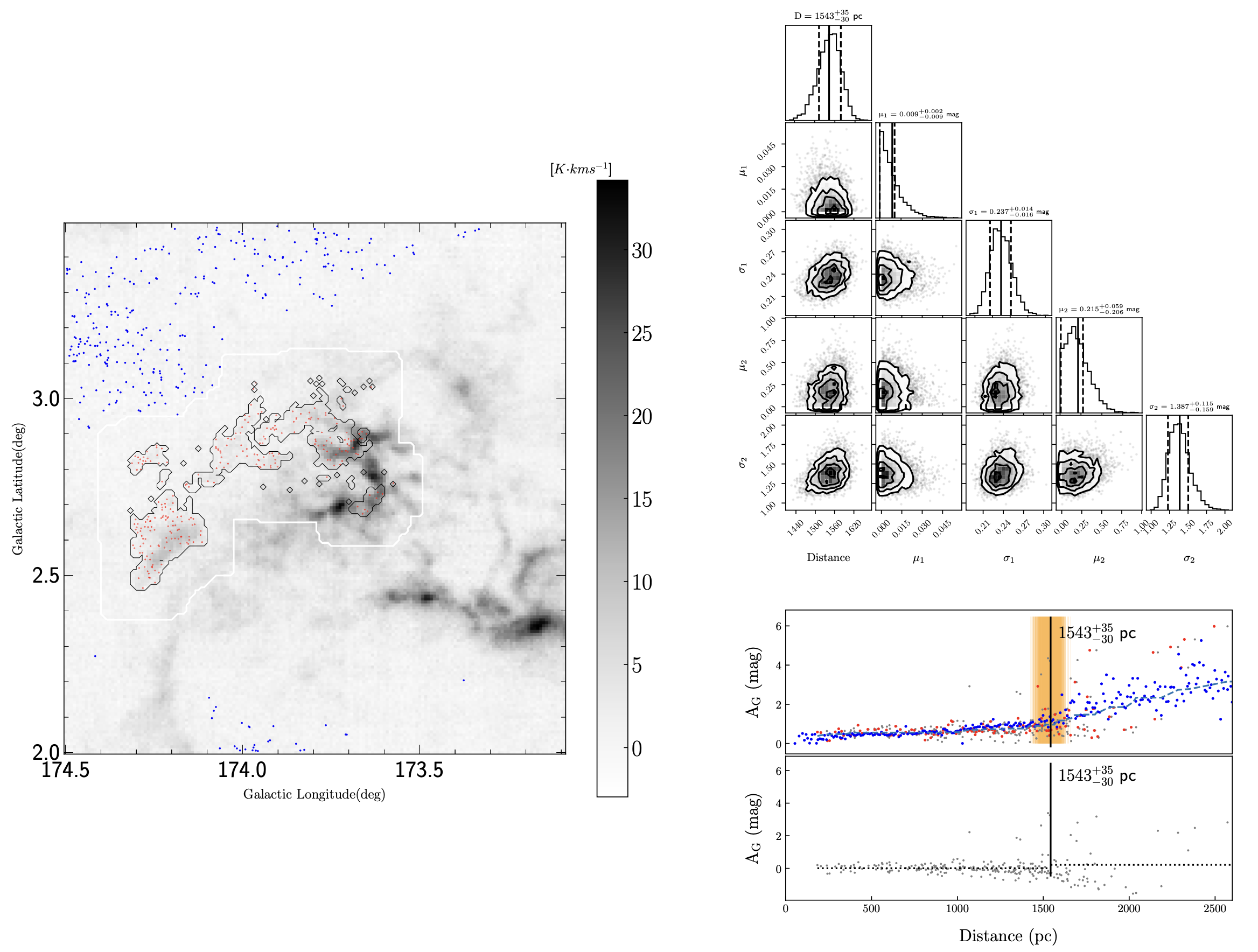}
    \caption{Distance measurement of molecular cloud component cloud-2 in S235 region, and mapping method is consistent with Figure \ref{fig:cloud1}. }
    \label{fig:cloud2}
\end{figure*}

\begin{figure*}
    \centering
    \includegraphics[width=\textwidth]{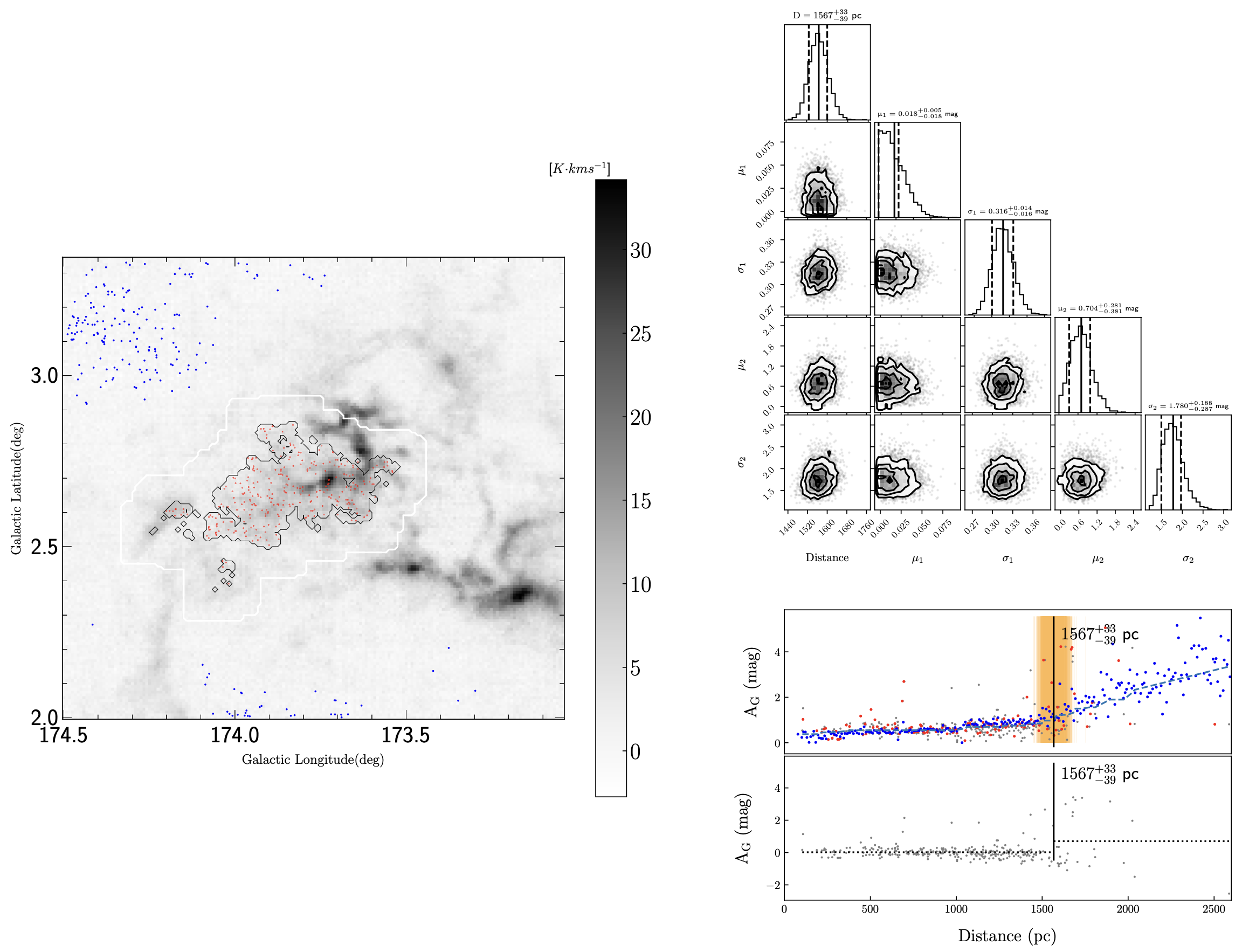}
    \caption{Distance measurement of molecular cloud component cloud-3 in S235 region, and mapping method is consistent with Figure \ref{fig:cloud1}.}
    \label{fig:cloud3}
\end{figure*}

The boundary of the cloud on the projection plane is used as the selection standard for on-cloud stars. 
For the reference area used for extinction background deduction, we use the integral intensity map of $^{12}$CO emission to select stars in an area with no or very weak $^{12}$CO emission. 
If the boundary of the cloud is indeed coherent in three dimensions and at a determinate distance, then all the stars within the boundary can result to the jump caused by the extinction of the cloud to the greatest extent. Different clouds, on the other hand, are likely to be located at different distances. 
According to the fitting results, for $^{13}$CO, most pixels have only one component on the spectral line, that is, there is no overlap between clouds at this position. 
Therefore, for a cloud with relatively large angular scale can be basically determined that the jump in star extinction is from this block structure. 
However, for most clouds, the overlap of light-of-sight direction is inevitable. 
We can analyze the measurement results as follows: \\
If two large cloud with partially overlap in the line-of-sight, firstly, in the non-overlapping part, we directly measure the distance of this part according to the extinction jump within the cloud boundary. 
However, in the overlapping part, we don't know yet how many stars are in front of, behind or between the two clouds. 
When we add these stars into our calculations, if the measured distances don't change much, then the stars are in front or behind the structure, which also means that the two clouds lay very close. 
On the other hand, if the measurement change significantly, such as showing multiple extinction jumps, then the respective distances between this component and its neighboring velocity component can be inferred by comparison. 
For example, we can firstly determine the distance of one cloud who has the maximum number of on-cloud stars. And then we can infer the distance of the other clouds by comparing to the relative extinction jumps.

According to the methods mentioned above, molecular clouds in the S235 region were decomposed into three components, and the distance parameters of each component were determined by the position of extinction jump. The measurement results were shown in Figure \ref{fig:cloud1}, Figure \ref{fig:cloud2} and Figure \ref{fig:cloud3} for component cloud-1, cloud-2 and cloud-3, respectively. And their physical properties are listed in table \ref{tab:cloud_dist}.

\begin{table*}
    \footnotesize
    \centering
    \caption{Properties of molecular cloud components of S235.}
    \begin{tabular}{ccccccccccc}
        %\hline
        \hline
       cloud id  & $l_{cen}$ & $b_{cen}$ & v$_{cen}$ & $\Delta v$ & Area & Flux & $\overline{N}$ & D & foreAg & backAg  \\
         & deg & deg & km s$^{-1}$ & km s$^{-1}$ & arcmin$^2$ & K km s$^{-1}$ & cm$^{-2}$ & pc & mag & mag \\
       \hline
       cloud-1  & 173.642 & 2.910 & -20.4 & 1.2 & 226.5 & 5144.5 & 5.5E+21 & 1547$^{+44}_{-43}$ & 0.09$\pm0.25$ & 0.408$\pm1.43$ \\
       cloud-2  & 173.937 & 2.773 & -19.8 & 1.8 & 310 & 4599.5 & 3.0E+21 & 1543$^{+35}_{-30}$ & 0.009$\pm0.237$ & 0.215$\pm1.387$ \\
       cloud-3  & 173.801 & 2.671 & -16.5 & 1 & 346.5 & 6677.9 & 4.1E+21 & 1567$^{+33}_{-39}$ & 0.018$\pm0.316$ & 0.704$\pm1.78$ \\
       \hline
    \end{tabular}
    \label{tab:cloud_dist}
\end{table*}

These three molecular cloud components basically cover the S235 region, as well as the aforementioned velocity extent. 
The cloud-1 has a central velocity of $-20.4$km s$^{-1}$, corresponding to a blue-shifted cloud associated with the S235-Main, while the cloud-3 has a central velocity of -16.5 km s$^{-1}$, corresponding to a red-shifted cloud associated with the S235-ABC. Moreover, the cloud-2 with the central velocity of -19.8 km s$^{-1}$ corresponds to the bridge feature of the intermediate velocity, who connects the cloud-1 and the cloud-3. 
Furthermore, the distances of these three molecular clouds are basically the same within 1$\sigma$ error range according to the distance measurement results, in which the cloud-1 is 1547$^{+44}_{-43}$ pc, the cloud-2 is 1543$^{+35}_{-30}$ pc, and the cloud-3 is 1567$^{+33}_{-39}$ pc, indicating that the cloud-1 and cloud-3 are likely to collide with each other at the same place, while the cloud-2 is the coherence bridge feature connecting the two clouds.

Based on the distances of these three clouds, we calculate the mean distance of the S235 molecular cloud complex to be 1552 pc, which is nearly 250 pc different from the 1.8kpc given in literature \citep[e.g.,][]{Evans1981, Dewangan2017a}, but close to the result of \cite{Kazes1977} of 1.6 kpc. Since our distance measurement is based on the parallax measurement method of Gaia satellite with higher accuracy, the distance of S235 is set as 1552 pc or 1.55 kpc hereafter in the paper.

\subsection{Clumps and cores \label{subsec:clump}}

Clumps and cores are fundamental units of star formation \citep[e.g.,][]{Bergin2007}. We searched for $^{13}$CO clumps because $^{13}$CO (J$=1-0$) is optical thin ($\tau\approx0.1\sim0.9$ with an average of 0.355, less than 1) thus can retain the properties of total molecular gas. We also checked the C$^{18}$O emission for each $^{13}$CO clumps in as much as it can trace higher density gas due to its lower abundance. On the other hand, we also identified the SCUBA-2 850 $\mu$m dust cores in the survey region relying on its higher sensitivity and small beam size (HPBW$=14^{\prime\prime}$, 1/4 of PMO-13.7m).  All the clumps and cores are obtained with $dendrograms$\footnote{\url{http://www.dendrograms.org/}} algorithm, which is able to extract hierarchical substructures of the dataset (2-D image or 3-D cube) into a set of iso-surfaces to form the skeleton with leaves or branches of the family tree. The top levels named 'leaves' of the tree are the most densest substructures of the dataset with a single local maximum, while the bottom levels named 'trunks' represent the low density extended structures. 

We performed 3-D (Position-Position-Velocity, PPV) dendrograms for $^{13}$CO and C$^{18}$O data cube. As for $^{13}$CO cube, we obtained clumps based on a minimum detection level of 2 $\sigma_{rms}$ (1 $\sigma_{rms}\sim$0.3 K) and then identified branches and leaves from their parent structures  with the parameters of the minimum increment of 2 $\sigma_{rms}$ and the minimum number of pixels of 54 (3 $\times$ 3 pixels $\times$ 6 velocity channels, corresponding to 1.6 beam size).  We also set a high threshold of 15 $\sigma_{rms}$ in order to obtain the densest structures and exclude the extended ones. In the paper, the data of C$^{18}$O is from the Milky Way Imaging Scroll Painting (MWISP) with rms level of about 0.3 K, which is not sensitive enough to detect the relative fainter C$^{18}$O clumps. In order to obtain them, we use a detection limit of 2 $\sigma_{rms}$. 
The minimum increment is also set to 2 $\sigma_{rms}$, and the same iso-surface pixel number of 54 is set. As a result, ten $^{13}$CO clumps were extracted by dendrograms and eight of them with C$^{18}$O clump counterparts. The physical properties of the $^{13}$CO clumps and the C$^{18}$O clumps are listed in Table \ref{tab:13clumps} and Table \ref{tab:18clumps}  respectively. Note that the poor emission of the C$^{18}$O may caused large uncertainties on mass estimation. 

The radius of the clumps were computed from the geometric mean of the intensity weighted second moments along the major and minor axis. 
And the mass of clumps were calculated from the standard LTE methods. We estimated the excitation temperature from the peak brightness temperature of $^{12}$CO due to its optically thick, and calculated the opacities, column densities and LTE masses of the clumps by accepting a constant [H$_2/^{12}$CO] abundance ratio of $1.1\times10^4$ \citep{Frerking1982} and isotopic ratio [$^{16}$O$/^{18}$O]$=560$ \citep{Wilson1994}, and [$^{12}$C$/^{13}$C]$=6.21d_{GC}+18.7$ which flows a gradient relation along the distance from the galactic center \citep{Milam2005}. At the distance $D=1.55$ kpc towards anti-galactic center, $d_{GC}$ of S235 is about 10.25 kpc thus [$^{12}$C$/^{13}$C]$=82$. Assuming that the clumps are spheres with radius R, we obtained the average H$_2$ densities and the virial parameters of the clumps. 

%%(analysis of the clump tables)

\begin{figure*}
\includegraphics[width=\textwidth]{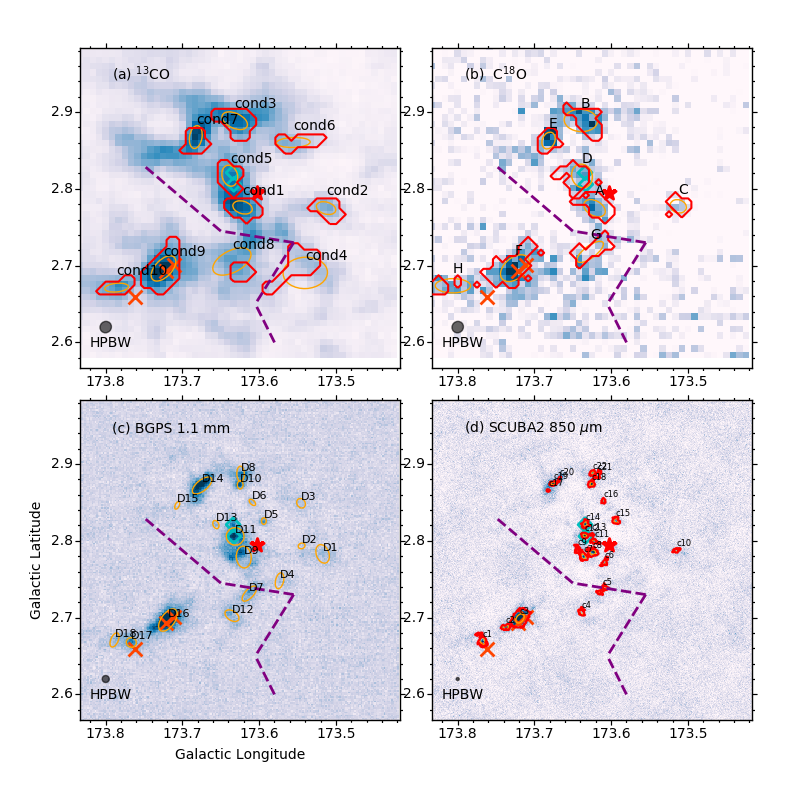}
\caption{Clumps and cores identification in the S235 complex. (a) the $^{13}$CO molecular clumps identified by using dendrogram. The red polygons indicate the dense leaves in dendrogram which were fitted by the orange ellipses. (b) the C$^{18}$O molecular clumps. (c) the BGPS dust clumps at 1.1 mm from \citet{Rosolowsky2010}. (d) the SCUBA-2 dust cores at 850 $\mu$m.  \label{fig:clump_ident}
}
\end{figure*}

2-D dendrograms in Position-Position plane were applied to SCUBA-2 850 $\mu$m image with setting the parameters of the minimum detection level to 2 $\sigma_{rms}$ (1 $\sigma_{rms}\sim0.06 Jy/beam$), the minimum increment to 2 $\sigma_{rms}$ and the minimum structure size to 9 pixels (3 $\times$ 3 pixels, corresponding to 0.86 beam size). And the minimum peak value is set to 3 $\sigma_{rms}$ to obtain the accurate leaves. As a result, a total of 22 dense cores were extracted from the image using the technique. All clumps and cores are presented in Figure \ref{fig:clump_ident}. 

Assuming a gas-to-dust ratio of 100, the beam-averaged H$_2$ column densities of the cores can be calculated from the integrated flux densities of the SCUBA-2 850 $\mu$m dust emission using the following formula:  

\begin{equation}
N(H_2)= \frac{100F_{\nu}}{\mu m_H\kappa_\nu B_\nu(T_d)\Omega_{beam}},
\end{equation}
where $F_\nu$ is the flux density, $\mu=2.8$ is the mean molecular weight, $m_H$ is the mass of a hydrogen atom, $\kappa_\nu$ represents the dust opacity per unit mass at 850 $\mu$m, which is interpolated from \cite{Ossenkopf1994} by $\kappa_\nu=\kappa_{230}(\frac{\nu}{230\ GHz})^\beta$, in which $\kappa_{230}=0.9\ cm^2g^{-1}$ is the opacity at 230 GHz and $\beta$ is the dust emissivity spectral index which is set to be 2, thus $\kappa_\nu$ at 850 $\mu$m is taken to be 0.21 $cm^2g^{-1}$, $B_\nu(T_d)$ is the Planck function, in which $T_d$ is dust temperature and is assumed to be 20 K, $\Omega_{beam}$ is the beam solid angle. The total mass of dust cores is derived from the integrated flux density over the target:

\begin{equation}
M_{dust}=\frac{100D^2S_{int}}{\kappa_\nu B_\nu(T_d)},
\end{equation}
where $D$ is the distance, and $S_{int}$ is the integrated 850 $\mu$m flux density. The physical properties of the identified dust cores are listed in Table \ref{tab:SCUBA2cores}. The core mass ranges from 5.5 to 31.1 M$_{\odot}$, with an extra core with extreme high mass of 269 M$_{\odot}$ (c3, which is located in the massive gas clump cond9). We also used the same method to calculate the beam-averaged column density and the dust clump mass for the 18 BGPS 1.1 mm dust clumps from \cite{Rosolowsky2010} found in the S235 region and then reproduced the properties catalog in Table \ref{tab:BGPSclumps}.

As mentioned above, the clumps and cores were identified with four tracers of $^{13}$CO, C$^{18}$O, BGPS 1.1 mm and SCUBA-2 850 $\mu$m, and each tracer represent different aspect of ISM, for example, optical thin $^{13}$CO can trace total molecular gas, and C$^{18}$O with low abundance can trace high dense gas, and long wavelength dust emission can trace dense gas as well. 
Since cores with smaller scale are substructure of clumps, hence, each clump can be resolved into several cores. 
The counterpart objects (i.e., C$^{18}$O clumps in Table \ref{tab:18clumps}, SCUBA-2 cores in Table \ref{tab:SCUBA2cores} and BGPS clumps in Table \ref{tab:BGPSclumps}) associated to the $^{13}$CO clumps are given in Table \ref{tab:associations}. 
Most of $^{13}$CO clumps hold $1\sim4$ cores, only two clumps with no core inside. 
There are four $^{13}$CO clumps namely cond5, cond7, cond9 and cond10 harbor Massive Young Stellar Objects (MYSOs),indicating massive star formation toward there.

%In Figure \ref{fig:mrrelation}, we studied the mass-size relationship for the $^{13}$CO, C$^{18}$O gas clumps, BGPS dust clumps and 850 $\mu$m dense cores found in the survey region. Mass-size relationship is a powerful tool to predict the tendency for a clump or core to evolve into a massive cluster, a massive star or a low mass star by a set of empirical bounds \citep[e.g.,][]{Kauffmann2010, Urquhart2013a}. As shown in Figure \ref{fig:mrrelation}, four $^{13}$CO clumps (cond1, 3, 7, 9), one C$^{18}$O clump (clump F), three BGPS dust clumps (D11, D14, D16) and one 850 $\mu$m dust core (c3) are found to have surface densities greater than the empirical lower bound for massive star formation, suggesting that they have potential  to form massive stars. The counterpart objects associated to the $^{13}$CO clumps are given in Table \ref{tab:associations}. Note that the poor emission of the C$^{18}$O may caused large uncertainties on mass estimation. One may notice that the cond9$, $clumpF$, $D16 and c3 represent the same site of the S235-AB which harbors Massive Young Stellar Objects (MYSOs), indicating that massive stars is being born toward there. 

\begin{table*}
\caption{Catalog of physical properties of $^{13}$CO molecular clumps in S235.}
\label{tab:13clumps}
\begin{threeparttable}
\begin{tabular}{ccccccccccc}
\hline
Clump & Glon & Glat & V$_{LSR}$ & $\Delta$V & R$_{eff}$ & Mass & $\alpha_{vir}$ & n & $\overline{N}$ & C$^{18}$O\tnote{a} \\
seq & $^{\circ}$ & $^{\circ}$ & km s$^{-1}$ & km s$^{-1}$ & pc & M$_{\odot}$ & &$\times10^3$cm$^{-3}$ & $\times10^{22}$cm$^{-2}$ & clump \\
\hline
  cond1 & 173.622 & 2.776 & -21.46 & 1.04 & 0.52 & 423.5 & 0.2 & 10.0 & 5.2 & A\\
  cond2 & 173.513 & 2.775 & -21.47 & 0.7 & 0.51 & 219.8 & 0.2 & 5.6 & 3.4 & C\\
  cond3 & 173.632 & 2.889 & -21.02 & 0.93 & 0.67 & 638.7 & 0.1 & 7.3 & 5.4 & B\\
  cond4 & 173.54 & 2.691 & -20.22 & 1.0 & 1.24 & 607.4 & 0.3 & 1.1 & 2.4 & -\\
  cond5 & 173.638 & 2.817 & -20.01 & 0.89 & 0.6 & 399.3 & 0.2 & 6.3 & 4.3 & D\\
  cond6 & 173.555 & 2.86 & -20.37 & 0.42 & 0.61 & 107.1 & 0.1 & 1.6 & 1.4 & -\\
  cond7 & 173.683 & 2.867 & -19.17 & 1.22 & 0.56 & 525.2 & 0.2 & 10.0 & 5.6 & E\\
  cond8 & 173.636 & 2.705 & -16.85 & 1.75 & 1.0 & 496.9 & 0.9 & 1.7 & 2.4 & G\\
  cond9 & 173.725 & 2.696 & -16.78 & 1.18 & 0.76 & 948.7 & 0.2 & 7.4 & 5.4 & F\\
  cond10 & 173.786 & 2.671 & -16.55 & 0.7 & 0.48 & 213.1 & 0.2 & 6.7 & 3.3 & H\\
\hline
%\tablecomments{$^{(a)}$Labels of associated C$^{18}$O clumps}
\end{tabular}
\begin{tablenotes}
    \footnotesize
    \item[a] Labels of associated C$^{18}$O clumps. 
\end{tablenotes}
\end{threeparttable}
\end{table*}

\begin{table*}
\caption{Catalog of physical properties of C$^{18}$O molecular clumps in S235. }
\label{tab:18clumps}
\begin{threeparttable}
\begin{tabular}{ccccccccccc}
\hline
Clump & Glon & Glat & V$_{LSR}$ & $\Delta$V & R$_{eff}$ & Mass & $\alpha_{vir}$ & n & $\overline{N}$ &${\rm N}_{\rm MYSO}$\tnote{a}\\
seq & $^{\circ}$ & $^{\circ}$ & km s$^{-1}$ & km s$^{-1}$ & pc & M$_{\odot}$ & &$\times10^3$cm$^{-3}$ & $\times10^{22}$cm$^{-2}$ & \\
\hline
  A & 173.622 & 2.775 & -21.67 & 0.79 & 0.62 & 252.6 & 0.2 & 3.7 & 2.2 & 0\\
  B & 173.639 & 2.889 & -21.01 & 1.15 & 0.9 & 510.7 & 0.35 & 2.4 & 2.4 & 0\\
  C & 173.513 & 2.777 & -21.54 & 0.62 & 0.49 & 103.9 & 0.21 & 3.1 & 1.6 & 0\\
  D & 173.638 & 2.817 & -19.98 & 1.09 & 0.73 & 251.6 & 0.51 & 2.2 & 1.6 & 2\\
  E & 173.682 & 2.863 & -19.06 & 0.94 & 0.49 & 208.1 & 0.29 & 6. & 2.2 & 1\\
  F & 173.725 & 2.697 & -16.8 & 1.39 & 0.88 & 612.5 & 0.44 & 3.1 & 2.5 & 2\\
  G & 173.627 & 2.718 & -17.15 & 0.77 & 0.64 & 153.2 & 0.33 & 2. & 1.6 & 0\\
  H & 173.806 & 2.674 & -16.29 & 1.25 & 0.74 & 298.1 & 0.6 & 2.5 & 2.2 & 1\\
\hline
%\tablecomments{$^{(a)}$Numbers of associated Massive Young Stellar Object (MYSO) candidates from \cite{Lumsden2013}.}
\end{tabular}
\begin{tablenotes}
    \footnotesize
    \item[a] Numbers of associated Massive Young Stellar Object (MYSO) candidates from \cite{Lumsden2013}.
\end{tablenotes}
\end{threeparttable}
\end{table*}

\begin{table*}
\caption{Catalog of physical properties of SCUBA-2 850 $\mu$m dust cores in S235. }
\label{tab:SCUBA2cores}
\begin{tabular}{cccccccccc}
\hline
Core & Glon & Glat & a & b & PA & S$_{int}$ & R$_{eff}$ & Mass  & $\overline{N}$ \\
seq & $^{\circ}$ & $^{\circ}$ & $^{\prime\prime}$ & $^{\prime\prime}$ & $^{\circ}$ & Jy & pc & M$_{\odot}$ & $\times10^{22}$cm$^{-2}$\\
\hline
  c1 & 173.768 & 2.67 & 16.9 & 9.2 & 113 & 1.86 & 0.21 & 28.0 & 7.4\\
  c2 & 173.737 & 2.687 & 11.2 & 6.3 & -169 & 1.43 & 0.14 & 21.49 & 5.7\\
  c3 & 173.719 & 2.699 & 20.9 & 13.5 & 51 & 17.85 & 0.28 & 269.0 & 71.0\\
  c4 & 173.638 & 2.707 & 10.5 & 7.1 & 100 & 0.53 & 0.14 & 7.98 & 2.1\\
  c5 & 173.611 & 2.736 & 17.5 & 7.7 & -137 & 0.76 & 0.19 & 11.46 & 3.0\\
  c6 & 173.608 & 2.772 & 12.3 & 5.4 & 57 & 0.61 & 0.14 & 9.21 & 2.4\\
  c7 & 173.636 & 2.779 & 10.8 & 7.7 & 122 & 1.36 & 0.15 & 20.46 & 5.4\\
  c8 & 173.624 & 2.785 & 11.5 & 9.0 & 172 & 1.98 & 0.17 & 29.83 & 7.9\\
  c9 & 173.644 & 2.789 & 11.8 & 6.4 & 96 & 0.51 & 0.15 & 7.68 & 2.0\\
  c10 & 173.515 & 2.787 & 9.2 & 4.7 & -161 & 0.37 & 0.11 & 5.59 & 1.5\\
  c11 & 173.622 & 2.799 & 12.3 & 4.8 & 161 & 0.68 & 0.13 & 10.26 & 2.7\\
  c12 & 173.634 & 2.807 & 7.9 & 6.2 & 157 & 1.66 & 0.12 & 25.0 & 6.6\\
  c13 & 173.625 & 2.808 & 5.7 & 3.4 & 173 & 0.37 & 0.07 & 5.52 & 1.5\\
  c14 & 173.634 & 2.822 & 9.1 & 7.3 & 61 & 1.22 & 0.14 & 18.32 & 4.8\\
  c15 & 173.594 & 2.826 & 8.7 & 7.4 & 138 & 0.86 & 0.13 & 12.99 & 3.4\\
  c16 & 173.61 & 2.852 & 6.5 & 4.7 & 89 & 0.36 & 0.09 & 5.43 & 1.4\\
  c17 & 173.682 & 2.865 & 3.8 & 3.0 & 177 & 0.39 & 0.06 & 5.9 & 1.6\\
  c18 & 173.625 & 2.874 & 8.1 & 6.7 & 66 & 1.5 & 0.12 & 22.6 & 6.0\\
  c19 & 173.675 & 2.875 & 10.1 & 4.9 & -166 & 2.07 & 0.12 & 31.15 & 8.2\\
  c20 & 173.667 & 2.88 & 7.7 & 2.5 & 67 & 0.64 & 0.07 & 9.68 & 2.6\\
  c21 & 173.617 & 2.886 & 11.4 & 6.0 & 50 & 0.85 & 0.14 & 12.79 & 3.4\\
  c22 & 173.624 & 2.888 & 7.7 & 5.5 & -143 & 0.77 & 0.11 & 11.68 & 3.1\\
\hline
\end{tabular}
\end{table*}

\begin{table*}
\caption{Catalog of physical properties of BGPS dust cores in S235. }
\label{tab:BGPSclumps}
\begin{tabular}{cccccccccc}
\hline
Core & Glon & Glat & a & b & PA & S$_{int}$ & R$_{eff}$ & Mass  & $\overline{N}$ \\
seq & $^{\circ}$ & $^{\circ}$ & $^{\prime\prime}$ & $^{\prime\prime}$ & $^{\circ}$ & Jy & pc & M$_{\odot}$ & $\times10^{22}$cm$^{-2}$ \\
\hline
  D1 & 173.517 & 2.783 & 45.08 & 29.59 & 70 & 1.34 & 0.32 & 58.2 & 2.7\\
  D2 & 173.545 & 2.793 & 16.42 & 12.97 & 159 & 0.19 & 0.13 & 8.4 & 0.4\\
  D3 & 173.545 & 2.849 & 23.47 & 19.09 & 58 & 0.44 & 0.18 & 19.0 & 0.9\\
  D4 & 173.573 & 2.748 & 38.02 & 18.81 & 103 & 0.55 & 0.23 & 23.8 & 1.1\\
  D5 & 173.594 & 2.826 & 15.44 & 11.19 & 91 & 0.48 & 0.11 & 20.9 & 1.0\\
  D6 & 173.609 & 2.85 & 18.69 & 10.01 & 44 & 0.22 & 0.12 & 9.4 & 0.4\\
  D7 & 173.613 & 2.731 & 44.98 & 19.94 & 133 & 1.95 & 0.26 & 84.3 & 4.0\\
  D8 & 173.624 & 2.887 & 35.62 & 17.57 & 89 & 2.09 & 0.22 & 90.4 & 4.2\\
  D9 & 173.62 & 2.779 & 51.09 & 34.89 & 85 & 5.13 & 0.37 & 222.0 & 10.0\\
  D10 & 173.625 & 2.873 & 19.33 & 13.6 & 91 & 1.04 & 0.14 & 45.1 & 2.1\\
  D11 & 173.631 & 2.806 & 41.67 & 40.84 & 17 & 5.98 & 0.36 & 258.8 & 12.0\\
  D12 & 173.635 & 2.703 & 36.4 & 22.23 & 33 & 1.43 & 0.25 & 61.9 & 2.9\\
  D13 & 173.656 & 2.821 & 20.0 & 13.36 & 66 & 0.26 & 0.14 & 11.0 & 0.5\\
  D14 & 173.674 & 2.872 & 54.88 & 25.34 & 141 & 6.77 & 0.33 & 293.0 & 14.0\\
  D15 & 173.707 & 2.846 & 19.37 & 9.34 & 116 & 0.17 & 0.12 & 7.2 & 0.3\\
  D16 & 173.719 & 2.697 & 60.35 & 31.63 & 126 & 15.15 & 0.38 & 656.0 & 31.0\\
  D17 & 173.767 & 2.668 & 23.77 & 22.38 & 24 & 1.62 & 0.2 & 70.2 & 3.3\\
  D18 & 173.789 & 2.671 & 35.29 & 17.39 & 113 & 0.99 & 0.22 & 42.9 & 2.0\\
\hline
\end{tabular}
\end{table*}

\begin{table*}
\caption{Catalog of objects associated to the $^{13}$CO clumps. }
\label{tab:associations}
\begin{threeparttable}
\begin{tabular}{ccccccccc}
\hline
Cond. & Glon & Glat & C$^{18}$O & BGPS & SCUBA-2 & ${\rm N}_{\rm MYSO}$ & MST YSO &Young star\tnote{a} \\
seq & $^{\circ}$ & $^{\circ}$ & clump & clump & core &  & group &cluster \\
\hline
  cond1 & 173.622 & 2.776 & A & D9 & c6, c7, c8, c9 & 0 & M5 & - \\
  cond2 & 173.513 & 2.775 & C & D1 & c10 & 0 & M2 & FSR 784\\
  cond3 & 173.632 & 2.889 & B & D8, D10 & c18, c21, c22 & 0 & M4 & S235 E2\\
  cond4 & 173.54 & 2.691 & - & D4 & -  & 0 & - & - \\
  cond5 & 173.638 & 2.817 & D & D11, D13 & c11, c12,c13, c14 & 2 & M3 & S235 Cl., CBB 2\\
  cond6 & 173.555 & 2.86 & - & D3 & - & 0 & - & - \\
  cond7 & 173.683 & 2.867 & E & D14, D15 & c17, c19, c20 & 1 & M1 & KKC 11\\
  cond8 & 173.636 & 2.705 & G & D7, D12 & c4, c5 & 0 & M6 & - \\
  cond9 & 173.725 & 2.696 & F & D16 & c2, c3 & 2 & M7 north & BDSB 72, 73, S235B Cl.\\
  cond10 & 173.786 & 2.671 & H & D17, D18 & c1 & 1 & M7 south & BDSB 71\\
\hline
%\tablecomments{$^{(a)}$Numbers of associated Massive Young Stellar Object (MYSO) candidates from \cite{Lumsden2013}.}
%\tablecomments{$^{(b)}$Names of associated young star clusters from \cite{Camargo2011}.}
\end{tabular}
\begin{tablenotes}
    \footnotesize
    \item[a] Names of associated young star clusters from \cite{Camargo2011}. 
\end{tablenotes}
\end{threeparttable}
\end{table*}

\iffalse
\begin{figure}
\includegraphics[width=\columnwidth]{Figures/Results/mosaic_L_clip_ddg_clumps_properties.eps}
\caption{Mass-size relationship of the clumps and cores identified in the S235 complex. The yellow shaded and green shaded regions demonstrate the empirical bounds of clumps without high-mass star formation \citep{Kauffmann2010} and clumps which are expected to evolve into young massive proto-clusters \citep{Bressert2012}. And the left region between them represents the clumps with high-mass star formation.  %Another criteria for massive star formation is that the surface densities of clumps should larger than 0.05 g cm$^{-3}$ \citep{Urquhart2013} but lower than 1 g cm$^{-3}$ \citep{Krumholz2008} which were shown as the lower and upper blue dotted lines, respectively. 
The upper and lower blue dotted lines represent the surface densities of 1 g cm$^{-3}$ \citep{Krumholz2008} and 0.05 g cm$^{-3}$ \citep{Urquhart2013}, which are upper and lower bounds of another empirical criteria of massive star formation. 
The blue solid circles indicate the $^{13}$CO molecular clumps while the skyblue circles indicate the C$^{18}$O molecular clumps. The magenta squares and red solid squares show the BGPS dust clumps and the SCUBA-2 dust cores, respectively. The magenta dashed line fits the mass-size relationship of the total clumps and cores which gives a slope of 1.94. %\textbf{(Dose it need to add error range?)}   
\label{fig:mrrelation}
}
\end{figure}
\fi

%\subsection{Temperature and density of the molecular cloud \label{subsec:temp}}

\subsection{Star formation in S235\label{subsec:sf}}

\subsubsection{Identification of young stellar objects\label{ysos}}
Low-mass young stellar objects (YSOs) can be classified as Class I, Class II and Class III objects according to their spectral indices \citep[e.g.,][]{Evans2009}. Different classes represent different evolutionary stages of YSOs, for Class I sources as the youngest protostars host circumstellar disks and infalling envelopes and Class II sources host optically thick disks, while Class III sources are the diskless pre-main-sequence stars. The YSOs with dusty circumstellar disks will cause infrared excess on their infrared colors. Many authors \citep[e.g.,][]{Gutermuth2009, Koenig2012, Koenig2014}, therefore, used NIR and MIR color excess schemes to find YSO populations. In this study, we employed UKIDSS-GPS NIR data and $Spitzer$-IRAC/MIPSGAL MIR data for finding deeply embedded and faint YSO sources. Similar work \citep{Dewangan2016} has been done toward the S235 region. However, their survey region was only limited to the S235-Main region, in addition to the S235-ABC region. Thus we employed the systematic YSO identification methods toward the entire S235 region. And a more elaborative description of these methods is given below. 

%(data preparation, survey region?)

Five independent methods are employed to select YSOs and its candidates in the present work. The first method is only based on the $Spitzer$-IRAC data, and the details of criteria are given in \cite{Gutermuth2009} Phase 1 (hereafter the method 1). The method was applied to the sources that have detections in all four $Spitzer$-IRAC bands with high quality ($\sigma<0.2$mag). First of all, we removed the extragalactic sources such as active star-formation galaxies, as their strong PAH-feature emission will cause spurious color excess in 5.8 and 8.0 $\mu$m bandpasses \citep{Stern2005} thus can be eliminated from our sample. The other extragalactic source contamination comes from broad-line AGN which have consistent MIR colors with YSOs \citep{Stern2005}. We utilized the [4.5] vs. [4.5]$-$[8.0] color-magnitude diagram to flag and remove AGN candidates as described in \cite{Gutermuth2009}. In addition, we removed sources classified as knots of shock emission and PAH-contaminated apertures. 
%More details of the contamination exclusion can be found in Figure \ref{fig:contam} in Appendix.  
After removing the contaminant sources, using the [3.6]$-$[4.5] vs. [4.5]$-$[5.8] color-color diagram, Class I young stellar objects can be identified by the following criteria, 
\begin{equation}
    \begin{aligned} \relax
    [3.6] - [4.5] &> 0.7 \\
    [4.5] - [5.8] &> 0.7
    \end{aligned}
\end{equation}
On the other hand, for Class II sources, the [3.6]$-$[5.8] vs. [4.5]$-$[8.0] color-color diagram is used, and the criterion are,
\begin{equation}
    \begin{aligned} \relax
    [3.6] - [5.8] + \sigma &\leq \frac{0.14}{0.04} \times ([4.5] - [8.0] - \sigma -0.5) +0.5 \\
    [3.6] - [5.8] - \sigma &> 0.35 \\
    [4.5] - [8.0] - \sigma &> 0.5 \\
    [3.6] - [4.5] - \sigma &> 0.15
    \end{aligned}
\end{equation}
Where $\sigma$ corresponds to the error of each color.

\begin{figure*}
\includegraphics[width=\textwidth]{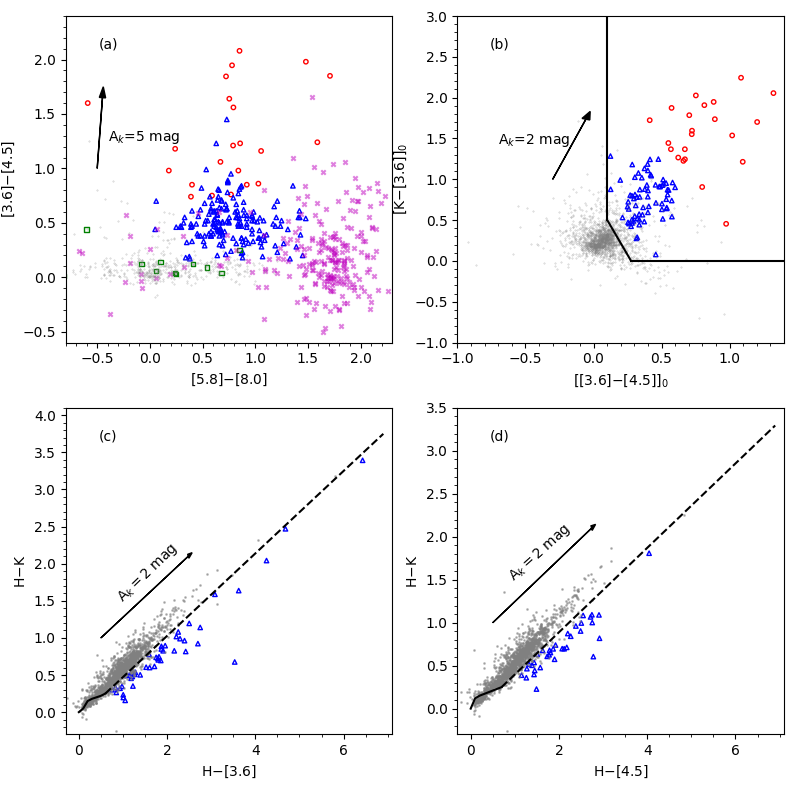}
\caption{YSOs identification schemes from method 1$-$4 (represent to a$-$d, respectively). (a) $Spitzer$-IRAC color-color diagram ([3.6]$-$[4.5] vs. [5.8]$-$[8.0]) \citep[from][Phase 1]{Gutermuth2009}. (b) The deredened color-color diagram ([K$-$[3.6]]$_0$ vs. [[3.6]$-$[4.5]]$_0$) using UKIDSS-GPS and $Spitzer$-IRAC data \citep[from][Phase 2]{Gutermuth2009}. (c) H$-$K vs. H$-$[3.6] color-color diagram \citep[from][]{Fang2012}. And the solid line represent the locus of the diskless stars. (d) Same with (c) but in H$-$K vs. H$-$[4.5] color-color diagram. For all panels, the red circles, blue triangles and green squares represent the Class I, Class II sources and Transition disks, respectively. The magenta crosses represent the contamination from PAH-rich galaxies, AGN galaxies, knots of shocked and PAH apertures. The arrows represent reddening and K-band extinction. 
}\label{fig:ysoclass}
\end{figure*}

As a result, we found 194 YSO candidates, of which 22 are Class I and 172 are Class II (see Figure \ref{fig:ysoclass}a.), and eliminated 378 PAH-rich galaxies, 400 AGN candidates, 1 knot of shock emission, and 266 PAH-contaminated apertures in the S235 region. 

In method 2, we considered the sources that lack detection on either 5.8 or 8.0 $\mu$m, but have high quality ($\sigma<0.1$mag) UKIDSS-GPS JHK bands (when J band is not available, H and K bands). The classification scheme was described in \cite{Gutermuth2009} Phase 2. The method utilizes the dereddened color-color space $[K-[3.6]]_0$ vs. $[[3.6]-[4.5]]_0$ to identify YSOs because line-of-sight extinction and intrinsic variability can cause a reddened Class II source be misclassified as a Class I source. The extinction was measured by baseline colors based on the Classical T Tauri Star (CTTS) locus of \cite{Meyer1997} and standard dwarf star colors \citep{Bessell1988} and the color excess ratios reported in \cite{Flaherty2007}. The measured colors were dereddened according to the reddening law by \cite{Flaherty2007}. With the above steps, the reddened Class II candidates can be separated from the Class I candidates. Finally, 22 Class I and 70 Class II sources were obtained from method 2  (see Figure \ref{fig:ysoclass}b.). 

After running Phase 1 and Phase 2 according to \cite{Gutermuth2009}, the next step (Phase 3) was to classify deeply embedded protostars and transition disks for the sources having excess emission at 24 $\mu$m. Embedded protostars are Class I candidates with bright emission at MIR wavelengths but lack emission at NIR wavelengths because they are deeply embedded in dust, while transition disks are Class II sources with significant dusty disks thus also show bright emission at MIR wavelengths. The classification criteria were given in \cite{Gutermuth2009} Phase 3. With this step, 1 embedded protostar and 12 transition disk YSOs were found toward the S235 region. They were shown in Figure \ref{fig:ysoclass}a if they have the respective colors. 

Due to the detection sensitivity limitations of IRAC [5.8] and [8.0] bands and the strong nebula background emission, more low-mass and low-luminosity members may be missed. Therefore, We use H$-$K vs. H$-$[3.6] and H$-$K vs. H$-$[4.5] color-color maps to search for lower mass young star objects. Details of the criteria can be found in \cite{Fang2012}. Figure \ref{fig:ysoclass}c\&d show the color-color diagrams of the two methods (hereafter the mthod 3 and method 4), and the reddening vector is from \cite{Indebetouw2005}. The classification criterion is roughly parallel to the reddening vector, and the stars above the dashed line of the criterion are diskless stars, most of which are unrelated foreground or background stars. The stars below the dashed line of the criterion have very large H$-$[3.6] or H$-$[4.5] color, which can not be explained by the reddening of diskless stars, indicating that these stars are young stellar object with disks, and they are simply classified into Class II YSOs. Using these two methods, we obtained 46 and 36 Class II sources respectively. 

\begin{figure*}
\centering
\includegraphics[width=0.5\textwidth]{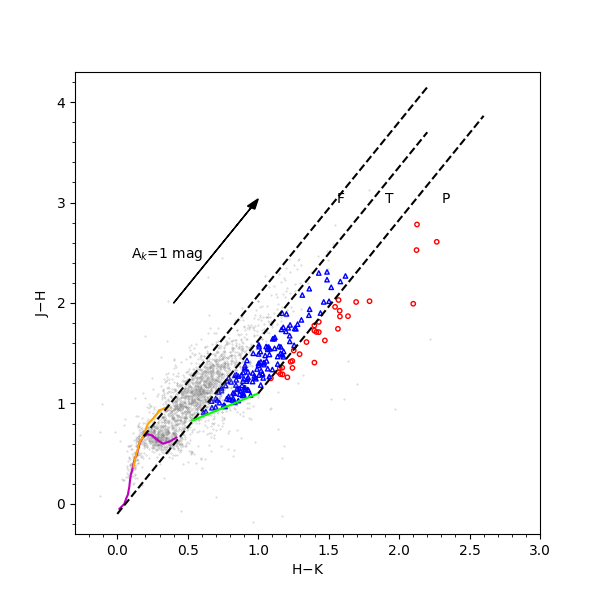}\includegraphics[width=0.5\textwidth]{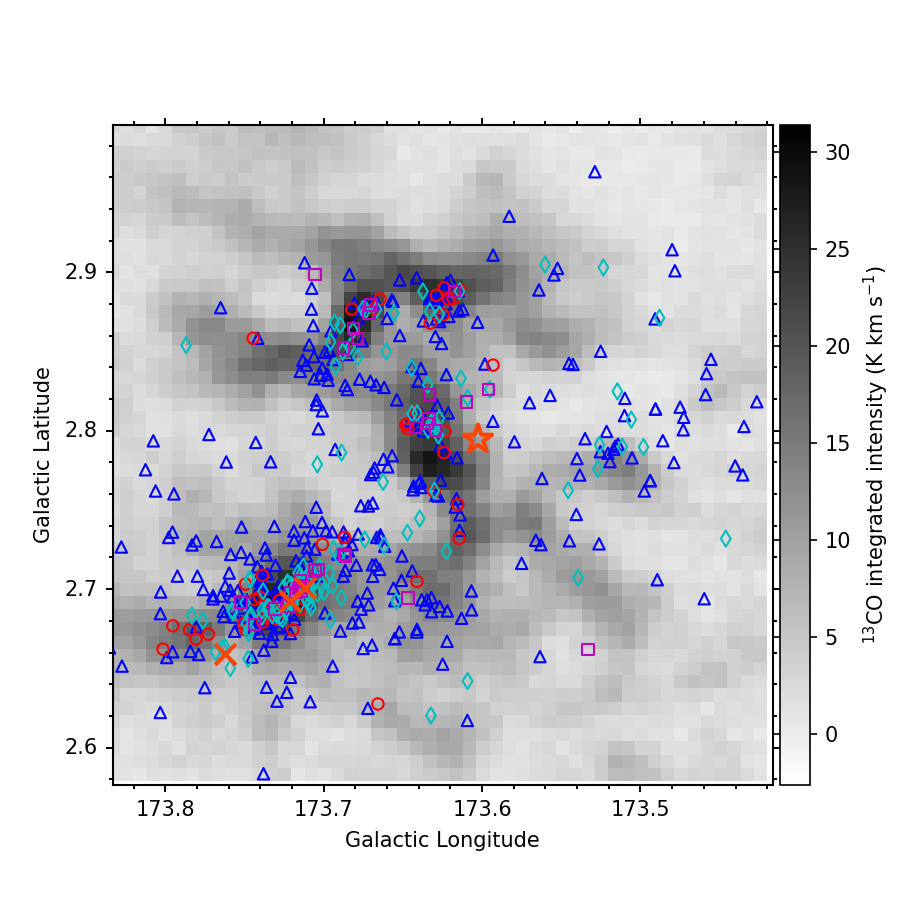}
\caption{Left panel: JHK Young Stellar Object (YSO) candidates identification scheme from method 5. The solid curves represent the unredened locus of main-sequence stars (magenta) and giants (orange) from \protect\cite{Bessell1988}, and Classical T Tauri (CTTS) stars (green) from \protect\cite{Meyer1997}. The extinction vector is drawn from \protect\cite{Indebetouw2005} extinction laws. The red circles represent Class I YSO candidates while the blue triangles represent Class II YSO candidates. Right panel: Spatial distribution of all YSOs and its candidates, with $^{13}$CO integrated intensity map in the background. The red circles represent Class I YSOs and the blue triangle represents Class II YSOs from menthod 1 to 4. The magenta squares represent Class I YSO candidates while the cyan diamonds represent Class II YSO candidates from method 5.} 
\label{fig:JHKyso}
\end{figure*}

Because UKIDSS-GPS is much more sensitive than $Spitzer$-IRAC, there are many faint YSOs ignored by IRAC but visible on UKIDSS. Therefore, we further use the UKIDSS-NIR color-color diagram (H$-$K vs. J$-$H) to search these sources. Note that the YSOs found by this method are not particularly reliable, because the SED of YSO has not reached its peak on this band, and its infrared color excess cannot very well reflect the properties of YSO disk. Therefore, we only take the stars identified by this method as the candidates for YSO. The NIR YSOs classification method was described in \cite{Sugitani2002} and \cite{Dewangan2016} (hereafter the method 5). They divided the color-color diagram into three different regions, namely "P", "T" and "F", according to the reddening bands of the Classical T Tauri (CTTS) locus \citep{Meyer1997} and the main-sequence and giants locus \citep{Bessell1988}. The left panel of Figure \ref{fig:JHKyso} shows the color-color diagram of YSO candidates identified by this method. The reddening vectors are calculated from the extinction law of \cite{Indebetouw2005}. "P" region represents the protostellar dominated region where the sources distributed in this region may be classified as Class I sources. "T" sources are located in the reddening bands of the CTTS locus thus were classified as Class II sources. "F" sources are falling between the reddening bands of the main-sequence and giant stars locus, which are considered to be field stars. By employing method 5, we obtained 32 Class I and 137 Class II sources. 

%However, the method 3 only considers the sources having detections in all JHK bands, there are still a significant number of sources that have detections only in the H and K bands. Therefore, many deeply embedded infrared excess sources were omitted by method 3. The color-magnitude (H-K vs. K) diagram allows us to select such sources by applying a color cut at H$-$K$>$1.04 \citep{Dewangan2016} (hereafter method 4). The color criterion was inferred from the color-magnitude analysis of the nearby control field. Finally, we obtained 132 additional deeply embedded sources (classified as Class II sources) using such simple cut. 

As a result of the five methods mentioned above, we totally found 550 YSOs and candidates, of which 77 are Class I and 473 are Class II sources (12 transition disks are included). And the spatial distribution of all YSOs is shown in the right panel of Figure \ref{fig:JHKyso}. It is uncertain whether an individual YSO is associated with S235, since there is no distance information for it. However, in our YSO catalog, the sources classified as Class I are distributed over the dense molecular clumps, so most of them are likely to associated with S235. On the other hand, the sources classified as Class II are distributed, and could be contaminated by field stars or extragalactic objects. Therefore, we use the MST clustering method to obtain YSO groups that are associated with S235.

\subsubsection{MST clustering analysis of YSOs}
A study of nearby cloud complexes indicates that young embedded clusters account for a significant (70-90\%) fraction of all stars formed in GMCs \citep{Lada2003}. The S235 complex, as a part of GMC G174+2.5, was found to host at least 9 embedded star clusters \citep{Camargo2011}. Clustered YSOs as the newborn members of young embedded star clusters thus are a good tracer of active star formation subregions in the molecular cloud. We utilized the Minimal Spanning Tree (MST) clustering method \citep[e.g.,][]{Barrow1985} to analyze the star formation substructure in the S235 region, such as groups/clusters of YSOs \citep{Gutermuth2009, Saral2015, Saral2017a, Sun2022}. MST is a unique network of a set of points (YSOs), in which the total length of the network is minimized and without closed loops. The clustering method extracts YSO groups/clusters by pruning an MST with a maximun cutoff distance ($d_c$) and a minimum number of members ($N$) in a group. 

There is no robust way to determine a cutoff distance for separating MST groups from "noise" YSOs, because the surface density profiles, fractions of members of YSOs, incompleteness of YSOs identification and heliocentric distance uncertainties vary from regions to regions. The simplest way is to directly use the average length $<l>$ of MST branches as the cutoff distance (i.e., 1 $\sim$ 3 times of  $<l>$, as did in \citealt{Barrow1985}). \iffalse {\scriptsize (Another commonly used method is the Path Linkage Criterion or PLC \citep{Battinelli1991a, Allen2008}. )}\fi Although this method has its advantage in detecting YSO groups, the thresholds they adopted are arbitrary. \cite{Gutermuth2009} constructed the Straight Line Fit (SLF) method to identify the local surface density threshold. The SLF method use straight lines to fit the short and long branch length domain of the cumulative distribution function of the MST branch lengths, and then choose the point of intersection as the threshold distance. More details were described in \cite{Gutermuth2009} and \cite{Saral2015}.

\begin{figure*}
\includegraphics[width=\textwidth]{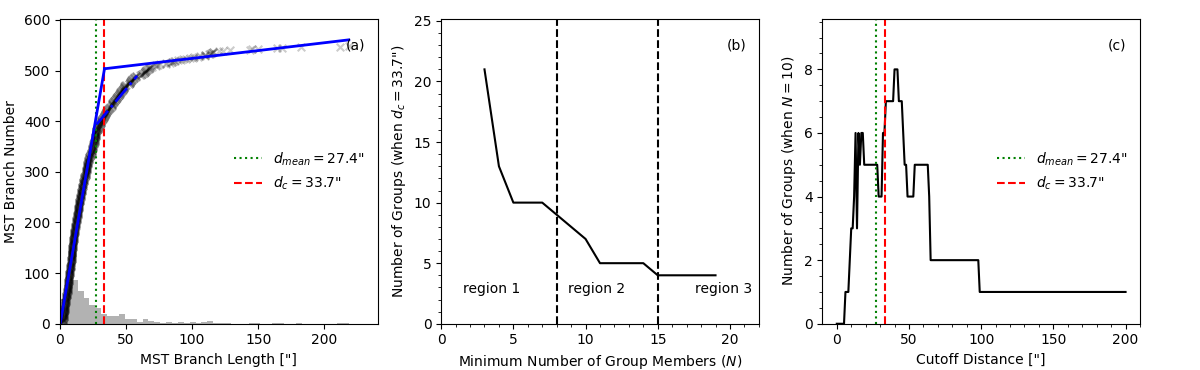}
\caption{MST clustering analysis of YSOs. (a) The SLF method \citep[from][]{Gutermuth2009}. The black solid line and crosses indicate the cumulative distribution function of the MST branch lengths, while the two blue solid lines fit the shot and long branch length domain, respectively. The red dashed line represents the intersection of the two fitting lines, giving the cutoff distance of $d_c=33.7^{\prime\prime}$. The green dotted line represents the average branch length. (b) The number of identified groups change with the minimum number of group members. (c) The number of identified groups changes with the separating distance.  \label{fig:mstana}}
\end{figure*}

Using the methods explained above, we performed the MST clustering analysis toward the sample of 550 YSOs we identified in S235. We first determined a cutoff distance ($d_c$) of MST branch lengths by employing the SLF method (see Figure \ref{fig:mstana}a.), and then used a standard value of $N=10$ as the minimum number of group members to identify YSO groups/clusters. 
In order to investigate how the number of identified groups changes with the minimum number of group members and the cutoff distances, we repeated the analysis with a set of test minimum numbers of group members in a range of $N=3 \sim 18$ when $d_c=33.7^{\prime\prime}$ (see Figure \ref{fig:mstana}b.), and a set of test cutoff distances in a range of $d_c=1^{\prime\prime} \sim 200^{\prime\prime}$ in steps of $1^{\prime\prime}$ when $N=10$ (see Figure \ref{fig:mstana}c.), respectively.

\begin{table}
\caption{Properties of the MST YSO groups.}
\label{tab:mstprop}
\begin{tabular}{cccccccc}
\hline
Name & Glon & Glat & N$_{IR}$ & I & II & II/I & Diameter  \\
  & $^{\circ}$ & $^{\circ}$ &   &   &   &   & $^{\prime\prime}$ (pc) \\
\hline
M1 & 	173.684 &	2.861 &	45  & 8 & 37 & 4.63 &  158 (1.19)\\
M2 & 	173.518 &	2.786 &	10 & 0 & 10 & - & 58 (0.44)\\
M3 &	173.635 &	2.808 &	38 & 11 & 27 & 2.45 &  128 (0.96)\\
M4 &	173.624 &	2.880 &	31 & 8 & 23 & 2.88 & 104 (0.78)\\
M5 &	173.635 &	2.764 &	10 & 1 & 9 & 9.00  & 44 (0.33)\\
M6 &	173.641 &	2.695 &	14 & 2 & 12 & 6.00  & 84 (0.63)\\
M7 & 	173.724 &	2.700 &	201 & 33 & 168 & 5.10 &  296 (2.22)\\
\hline
\end{tabular}
\end{table}

In Figure \ref{fig:mstana}a, the intersection of the two straight fitting lines indicates a cutoff distance of $33.7^{\prime\prime}$ ($\sim 0.25$ pc at the distance of 1.55 kpc). When involved cutoff distance into the MST analysis, the number of identified MST groups monotonously decreases with the minimum number of group members as shown in Figure \ref{fig:mstana}b. Such distribution can be separated into three regions. In region 1 ($3 \leq N \leq 8$), the numbers of identified groups rapidly decrease as the number of group members slightly increase, which suggests that small groups are significantly affected by contamination which accidentally have similar number of group members as small groups. In region 2 ($8 \leq N \leq 15$), the number of identified groups slightly decreases from nine groups to four groups, suggesting that groups or subgroups can be identified within this region, and the smallest group should contain at least 8 group members. Within this region, the number of groups keep steady and seven groups were identified with $N=10$, suggesting that the standard value of $N=10$ is good to identify substructures of YSO distribution in our study. In region 3 ($N\geq15$), four large groups remain unchanged. Given the standard value of $N=10$ in the MST analysis (Figure \ref{fig:mstana}c), the number of groups change changed as a crescent function with a set of cutoff distances and finally 7 groups were identified with the cutoff distance of $d_c=33.7.0^{\prime\prime}$. Note that the identified group number reaches its maximum of 8 groups/subgroups when setting the cutoff distance in a range of $40^{\prime\prime} \sim 48^{\prime\prime}$ which is longer than the cutoff distance given by SLF method.

\begin{figure*}
\includegraphics[width=\textwidth]{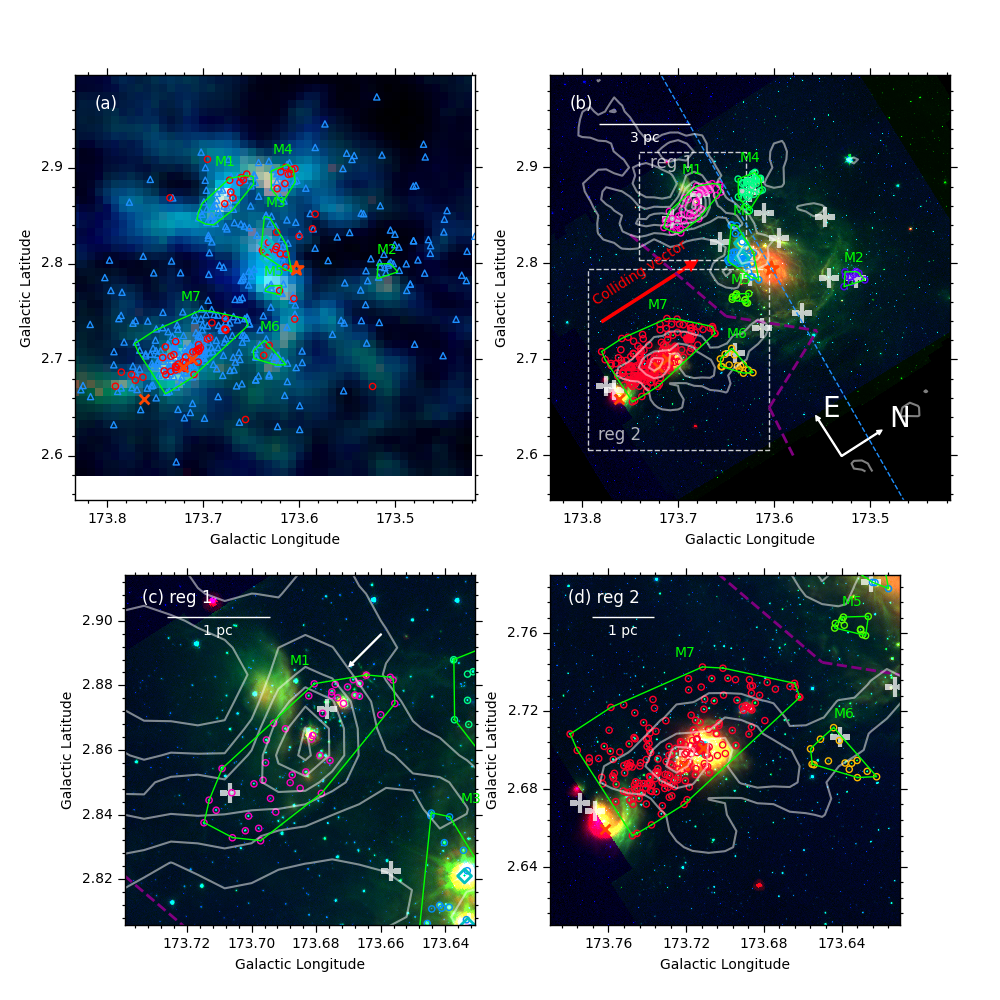}
\caption{Distribution of the MST YSO groups. (a) The distribution of all YSOs with MST groups. The background image is the CO color composite image, in which blue is $^{12}$CO, green is $^{13}$CO, and red is C$^{18}O$. The red circle represents Class I and the blue triangle represents Class II. (b) The color composite image is coding from MIPS 24 $\mu$m (Red), IRAC 4.5 $\mu$m (Green) and UKIDSS-GPS K band (Blue). The circles in different colors represent different YSO groups which are also highlighted by the green polygons. The white crosses indicate the BGPS 1.1 mm dust clumps. The white contours represent the $^{13}$CO distribution of the bridge feature (integrated from $-17.5$ $\sim$ $-19.5$ km s$^{-1}$). The contours start from 3 K km s$^{-1}$(3 $\sigma$) to 21 K km s$^{-1}$ in steps of 3 K km s$^{-1}$. The magenta dashed line represents the colliding interface found by the moment 2 map. The blue dashed line divides the S235 complex into two parts. The southern part was strongly affected by CCC process while the northern part was collision quiet. And the red arrow indicates the colliding vector from the south to the north. (c) and (d) The zoomed-in regions of M1 and M7 from (b).  \label{fig:yso_groups}}
\end{figure*}

To conclude, with both $N=10$ and $d_c=33.7^{\prime\prime}$ parameters involved in our MST analysis, we identified 7 MST groups which contains 63$\%$(349/550) of the total numbers of YSOs in the S235 region. The MST groups consist of 10 $-$ 201 YSOs with diameters of 0.33 $-$ 2.22 pc. The spatial distribution of the MST groups is shown in Figure \ref{fig:yso_groups} and their properties are tabulated in Table \ref{tab:mstprop}. 

As shown in Figure \ref{fig:yso_groups}, the YSOs are clustering toward the star formation subregions, and most of them are still embedded in molecular gas. The spatial distribution of the bridge feature is drawn in white contours in the Figure \ref{fig:yso_groups}b\&c\&d, which represents the morphology of the collisional influenced gas. The bridge feature only distribute at the southern and southeastern part of the S235 complex as separated by the blue dashed line in the figure, suggesting that the star formation activities are seriously affected by the CCC process in the southern regions but quiet in the northern regions. Considering the colliding interface and the distribution of the bridge feature, the colliding vector of the two clouds should be pointing from the south to the north (depicted as the red arrow), which is perpendicular to the blue dashed line in Figure\ref{fig:yso_groups}b. There are at least five MST groups (M1, M3, M5, M6 and M7) associated with the bridge feature, suggesting that the star formation in these sites might be triggered by the CCC process. The zoomed-in regions of these YSO groups is shown in Figure \ref{fig:yso_groups}c\&d. The YSO group M1 (namely the YSO Wall) and M7 show a elongated shape with a length of $\sim$ 2 pc and $\sim$ 3 pc, respectively. On the other hand, the bridge feature distribution also shows a close shape to the YSO group, indicating that the YSOs (at least the Class I YSOs) in these MST groups (or young star clusters) may form from the dense gas compressed by CCC process.

\subsection{The effect of cloud-cloud collision on star formation} \label{sec:discussion}

%\subsection{Star and core formation efficiency\label{subsec:sfe}}

Our systematically census of young stellar objects suggested that the S235 complex is a very active star formation region with a width mass range of young stellar populations clustering in seven subregions. we obtained the core formation efficiency (CFE), the star formation efficiency (SFE) and rate (SFR) by counting the number of dust cores and YSOs in each $^{13}$CO molecular clump. The CFE or SFE is defined as below, 
\begin{equation}
{\rm CFE}\, ({\rm SFE}) = \frac{\rm M_x}{\rm M_x + M_{cloud}},
\end{equation}
where ${\rm M_x}$ is the total mass of cores (CFE) or stars (SFE) and ${\rm M_{cloud}}$ is the total cloud mass. CFE and SFE indicate the fraction of molecular gas evolved into dense cores and stars. SFE varies from less than 0.1\% to 50\% among molecular clouds with a median value of 2\% in the inner Galaxy \citep[][]{Myers1986}.

As mentioned earlier, only the Class I and Class II YSO candidates were identified by our NIR-MIR YSO selection scheme, so that we could not calculate the total stellar mass of each $^{13}$CO clump directly due to the lack of information on the number of the evolved stars (Class III and main-sequence stars) and the mass of each star. Therefore, we made the following assumptions in calculation: (1) A disk fraction ($N$(II)/$N$(II$+$III)) is considered to be 50\% \citep[][]{Fang2013, Gong2016}, given $N$(III) $=N$(II). (2) The mass distribution of YSOs follows the initial mass function (IMF) of \cite{Chabrier2003}. (3) Accept a lower magnitude limit of the identified YSOs in K band to 14 mag, which corresponds to 3.05 absolute magnitude at the distance of 1.55 kpc. This gives the lower mass limit of 0.03 M$_{\odot}$ \citep[][]{Siess2000}. We also simply assumed the upper limit of the IMF to be 80 M$_{\odot}$ \citep[][]{Gong2016}, because the number of massive stars decrease seriously in IMF thus will not affect the total mass significantly. Taking these assumptions into account, one can calculate the total stellar mass by integrating the IMF. In addition, this result should also be added to the mass of the Massive Young Stellar Objects (MYSOs) as its final total stellar mass, and we assume that the average mass of each MYSO is 10 M$_{\odot}$. The calculation results are listed in Table \ref{tab:13cosfe}. To conclude, the average mass of YSOs is about 0.64 M$_{\odot}$ which is in agreement with observations. The CFE for each $^{13}$CO clump ranges from 2.5\% to 23.4\%,  while the SFE is in a range of 2.9\% $-$ 21.2\% which is greater than the SFE (3.0\%$-$ 6.3\%) in nearby clouds \citep[][]{Evans2009} and the median value (2\%) in the inner Galaxy \citep[][]{Myers1986}. Note that we only counted the YSOs who are the members of the MST groups as well as enclosed by the area of a clump, so that some clumps did not present SFE even though they harbors some scattered YSOs. The uncertainties of CFE and SFE are estimated by the error from gas mass (40\% of gas mass \citep[][]{Beaumont2013}) and the error from YSO mass (Poisson error $\sqrt{N}/N \times {\rm M_{\star}}$). %There should be a factor of 2 uncertainties when calculating the CFE and SFE due to the projection effect from light-of-sight. (?)

\begin{table*}
\caption{Catalog of derived properties of $^{13}$CO clumps in S235.}
\label{tab:13cosfe}
\begin{tabular}{ccccccccccccc}
\hline
Cond. & Glon & Glat & R$_{eff}$ & M$_{gas}$ & M$_{cores}$ & CFE
 & N$_{\rm YSOs}$ & N$_{\rm MYSOs}$ &M$_{stars}$ &SFE
 &SFR   \\
seq & $^{\circ}$ & $^{\circ}$ & pc & M$_{\odot}$ & M$_{\odot}$ & \% &I+II & &M$_{\odot}$ & \% &M$_{\odot}$/Myr \\
\hline
  cond1 & 173.622 & 2.776 & 0.52 & 423.5 & 67.2 & 13.7$\pm$6.7 & 0+10 & 0 & 12.7 & 2.9$\pm$1.3 & 6.4 \\
  cond2 & 173.513 & 2.775 & 0.51 & 219.8 & 5.6 & 2.5$\pm$1.4 & 0+10 & 0 & 12.7 & 5.5$\pm$2.4 & 6.4 \\
  cond3 & 173.632 & 2.889 & 0.67 & 638.7 & 47.1 & 6.9$\pm$3.6 & 8+23 & 0 & 34.4 & 5.1$\pm$2.0 & 17.2 \\
  cond4 & 173.54 & 2.691 & 1.24 & 607.4 & - & - &  0 & 0 & - & - &  -& \\
  cond5 & 173.638 & 2.817 & 0.6 & 399.3 & 59.1 & 12.9$\pm$6.4 & 11+27 & 2 & 61.4 & 13.3$\pm$4.8 & 30.7 \\
  cond6 & 173.555 & 2.86 & 0.61 & 107.1 & - & - & 0 & 0 & - & - &  -\\
  cond7 & 173.683 & 2.867 & 0.56 & 525.2 & 46.8 & 8.2$\pm$4.2 & 8+37 & 1 & 62.2 & 10.6$\pm$3.9  & 31.1\\
  cond8 & 173.636 & 2.705 & 1.0 & 496.9 & 19.5 & 3.8$\pm$2.1 & 2+12 & 0 & 16.6  & 3.2$\pm$1.4 & 8.3 \\
  cond9 & 173.725 & 2.696 & 0.76 & 948.7 & 290.5 & 23.4$\pm$10.2 & 33+168 & 2 & 255.0 & 21.2$\pm$6.7 & 127.5  \\
  cond10 & 173.786 & 2.671 & 0.48 & 213.1 & 28.0 & 11.6$\pm$5.8 & 5+13 & 1 & 29.7 & 12.2$\pm$4.7 & 14.9  \\
\hline
\end{tabular}
\end{table*}

After calculating the stellar mass toward each $^{13}$CO clump, we can directly derive the SFR by, 

\begin{equation}
{\rm SFR} = \frac{\rm M_{\star}} {\tau}, 
\end{equation}
where $\tau$ is the average age of YSOs assumed to be 2 Myr \citep[][]{Evans2009}. The SFRs varies from 6.4 to 127.5 M$_{\odot}$ Myr$^{-1}$ among the clumps. Three clumps (cond5, cond7 and cond9) with high values of SFR (30.7, 31.1 and 127.5 M$_{\odot}$ Myr$^{-1}$, respectively) are associated with the M3, M1 and M7 YSO groups, in which the star formation are possibly triggered by collision.

\begin{figure}
\includegraphics[width=\columnwidth]{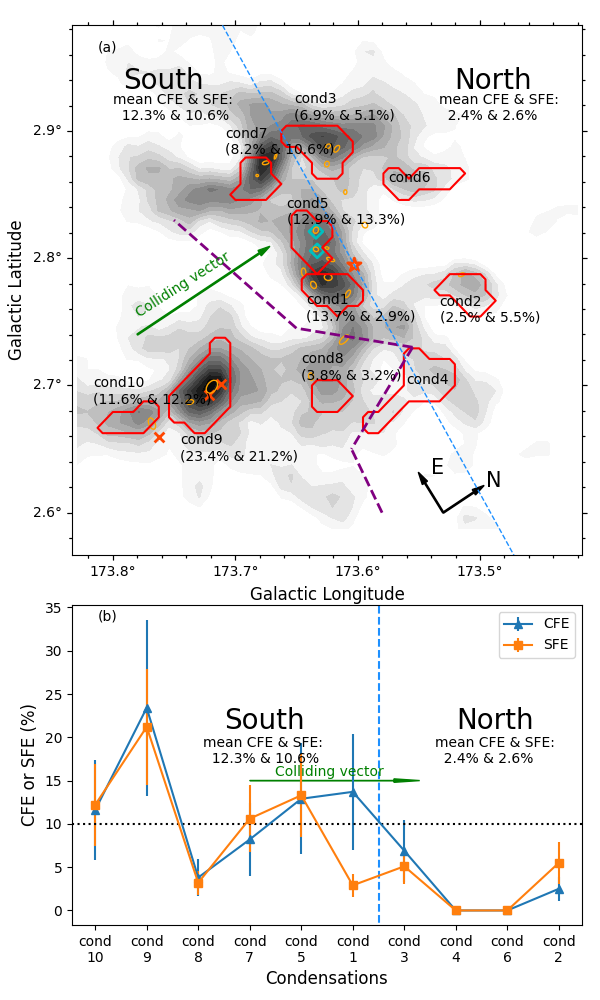}
\caption{(a) Spatial distribution of core and star formation efficiency. The gray scale image represents the $^{13}$CO data. The red polygons indicate the $^{13}$CO clumps while the orange ellipses represent the SCUBA-2 dust cores at 850 $\mu$m. The values of CFE and SFE for each clump are drawn in the figure, and the first and the second number represents the CFE and SFE, respectively. The blue dashed line divides the S235 complex into two subregions, the "South" and the "North", according to the distribution of the bridge feature in Figure \ref{fig:yso_groups}. The green arrow indicates the colliding vector which perpendicular to the blue dashed line. (b) The CFE and SFE profiles of the complex from south to north in terms of the $^{13}$CO clumps. The blue and orange solid line represents the profile of CFE and the SFE, respectively. The green arrow and the blue dashed line are drawn in the same manners as in (a). The mean CFE and SFE in the southern subregion (12.3\% \& 10.6\%) is about 3 $\sim$ 5 times than that of in the northern sub-region (2.4\% \& 2.6\%) of the complex. \label{fig:cfesfe}}
\end{figure}

When drawing the spatial distribution of CFE and SFE for the ten $^{13}$CO clumps (see Figure \ref{fig:cfesfe}a), the S235 complex can also be divided into two subregions (the South and the North) as it has been shown in Figure \ref{fig:yso_groups}b by a blue dashed line. We also plotted the CFE and SFE profiles of the complex along the colliding vectors from the south to the north in terms of the identified $^{13}$CO clumps (see Figure \ref{fig:cfesfe}b). 4/6 of the clumps in the southern subregion present higher CFE and SFE larger than 10\% with a mean value of 12.3\% and 10.6\%, respectively. However, all of the clumps in the northern subregion host lower CFE and SFE with an average value of 2.4\% and 2.6\%, respectively. Note that the star formation in the southern subregion is seriously affected by the CCC process, while the northern subregion is influenced by the expansion of the S235 \textsc{Hii} region, indicating that the CCC process can significantly enhance the CFE and SFE of the clouds by 3 $\sim$ 5 times larger than that of the clouds without collision influence.

%\subsection{Effects of triggered star formation\label{subsec:trigger}}

%High-mass stars strongly affect environment gas via ultroviolet (UV) photons, stellar winds, and supernova explosions. \cite{Elmegreen1977} proposed that the expanding ionization front plays a constructive role in inciting star formation activities. In generally, the "collect \& collapse" and "radiation driven implosion (RDI)" processes are thought to be promising pictures to explain the star formation at the periphery of an \textsc{Hii} region. However, both the two models have not been established due to the confrontation between the theroes and observations. In the "collect \& collapse" scenario, the ambient meterial accumulated by the expanding motions of the ionized gas to form a dense molecular layer, which becomes gravitationally unstable and fragments and collapses to form the next-generation of stars \citep{Elmegreen1977}. In the "radiation driven implosion (RDI)" scenario, a pre-existing clump is further compressed by the photoionization induced shock, and then collapse to form stars (Bertoldi 1989, Lefloch \& Lazareff 1995). It was hard to distinct these two effect of triggering star formation in observations because we only have a "snapshot" view toward an onset star formation region. 

%\subsection{CCC scenario in S235 \label{subsec:ccc}}

\begin{figure}
\includegraphics[width=\columnwidth]{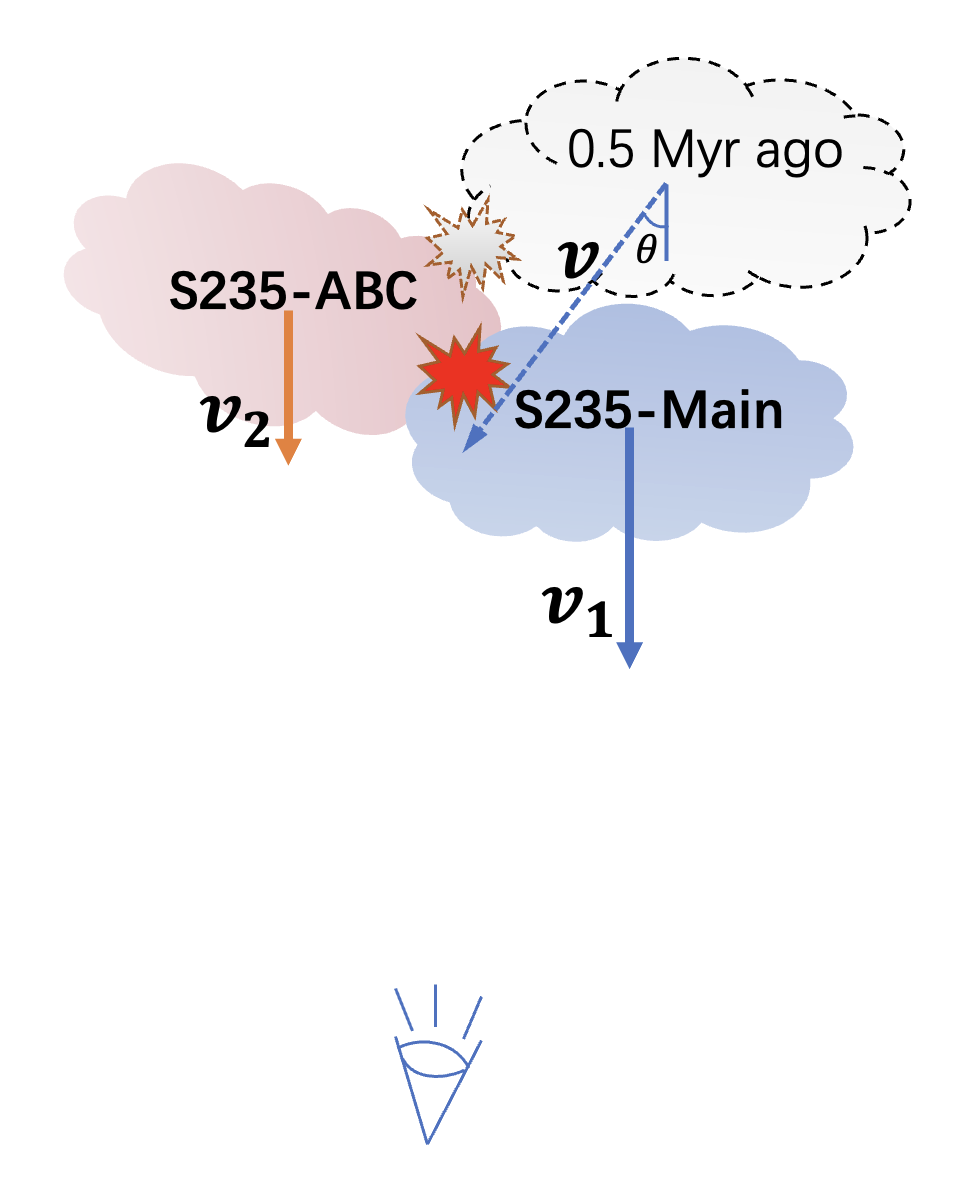}
\caption{Cloud-cloud collision scenario in S235. The S235-Main is much closer and approaching us at a much faster speed (4 km s$^{-1}$relative to S235-ABC), indicate that the two molecular clouds are probably moving out of the collision state. However, both the massive young stellar objects and Class I YSOs formed before 0.5 Myr, which is the most intense moment of cloud-cloud collision. At this point, the S235-Main collided with the S235-ABC with an incident angle of $\theta$ at a colliding velocity greater than 4 km s$^{-1}$. 
}\label{fig:scenario}
\end{figure}

As mentioned above, our present work supports that the cloud-cloud collision process might have taken place in the S235 complex. The most direct evidence of cloud-cloud collision is that the two clouds meet at the same position with a supersonic relative speed. S235 is consistent with this scenario, as shown in Figure \ref{fig:scenario}, the S235-Main (1547$^{+44}_{-43}$ pc) and S235-ABC (1567$^{+33}_{-39}$ pc) are nearly located at the same position within 1$\sigma$ error range, indicating that they are in contact with each other. On the other hand, the S235-Main is approaching us at a blue-shifted velocity of $-$20.5 km s$^{-1}$, while the S235-ABC is at a slower blue-shifted velocity of $-$16.5 km s$^{-1}$, indicating that the cloud-cloud collision process has occurred in S235 or the two clouds are currently moving away from each other. 
This result is consistent with the observational signatures of cloud-cloud collision process, such as the broad bridge feature connecting the two clouds with a supersonic separating velocity, the distinct complementary distribution of the two clouds and the broad line widths of CO emission in the colliding interface. As did in \cite{Dewangan2017a}, they compared the typical collision time scale ($\sim$0.5 Myr) with the dynamical ages of the compact \textsc{Hii} regions in the S235-ABC (0.06$-$0.22 Myr) and the average ages of the YSOs (Class I of $\sim$0.44 Myr and Class II of 1$\sim$3 Myr), and found that the formation of the youngest populations and massive stars might be induced by the CCC about 0.5 Myr ago. The tight association between the YSO groups (M1 and M7) and the bridge feature (collisional impact gas) also supports the conclusion.

In summary, the picture of cloud-cloud collision process in the S235 molecular cloud complex can be inferred as follows: the S235-Main and the S235-ABC began to collide with each other in about $3\sim5$ Myr ago, triggering the formation of large amounts of low-mass, aged Class II YSOs. And then in about 0.5 Myr ago, more dense gas was compressed to form younger stellar population in groups (M1, M3, M5, M6, and M7) and massive young stellar objects (MYSOs) in M1, M3, and M7, as well as the O-type star in the S235-Main and B-type stars in S235-ABC. 
Finally, to the present day, the two molecular clouds may pull out of collision state and moving away from each other. 
In order to form massive objects, a higher CFE or SFE is required. The CCC process as a mechanism can enhance the CFE and SFE to over 10\%, which is significantly (3$\sim$5 times) larger than the clouds whose star formation was triggered by the expansion of \textsc{Hii} region or the clouds without collision influence. 
%Our work expands their outcomes on the relationship between the cores/YSOs and the CCC effects. 

\section{Summary} \label{sec:conclusions}

In this paper, we have performed a study of cloud-cloud collision (CCC) toward the S235 complex and its effects on triggering star formation using the CO data from PMO-13.7m telescope and the archival data. The main results are summarized as follows.

\begin{enumerate}

\item  we confirmed the results the cloud-cloud collision process in S235 with the observational signatures of the supersonic velocity separation ($\sim$4 km s$^{-1}$), the broad bridge feature, the colliding interface (with large velocity dispersion of $\sim$2 km s$^{-1}$) and the distinct complementary distribution between the S235-Main and the S235-ABC. 
\item The high precision Gaia parallax measurement method was used to determine the distance of two main molecular cloud components in S235, the S235-main and the S235-ABC, which is in a distance of 1547$^{+44}_{-43}$ pc and 1567$^{+33}_{-39}$ pc, respectively. And their relative distance is less than the error range of 1 $\sigma$, indicating that these two clouds are probably colliding with each other at the same place. Meanwhile, we used their average distance of 1552 pc as the distance of S235. 
\item Ten $^{13}$CO clumps and 22 dense cores were identified using the PMO-13.7m $^{13}$CO data and SCUBA-2 data, respectively. And each clump hosts 1 $\sim$ 4 cores with mass ranges from 5 $-$ 270 M$_{\odot}$. 
\item A total of 550 YSOs were identified, and 63\% of them were clustering into 7 MST groups (M1$-$M7). The formation of YSOs groups (M1, M3, M5, M6 and M7), and the massive young stellar objects (MYSOs) in M1, M3 and M7 may be triggered by the CCC process, which enhance the CFE and SFE of cloud to over 10\%, which is 3$\sim$5 times larger than that without the influence of collision. 
%\item The S235-Main and the S235-ABC collided with each other in about $3\sim5$ Myr ago and produced large amounts of compressed gas to form massive young stars in about 0.5 Myr ago. Until today, the two molecular clouds may be out of collision and moving away from each other. 

\end{enumerate}

\section*{Acknowledgements}

We acknowledge support by the National Natural Science Foundation of China (NSFC, Grant No. 12041305, 12373026). 
This work is based on data obtained as part of the Milky Way Imaging Scroll Painting (MWISP) science project conducted by Purple Mountain Observatory (PMO). 
This paper made use of data products from the UKIRT Infrared Deep Sky Survey, the Two Micron All Sky Survey (a joint project of the University of Massachusetts and the Infrared Processing and Analysis Center/California Institute of Technology, funded by NASA and NSF) and the archived data obtained with the $Spitzer$ Space Telescope (operated by the Jet Propulsion Laboratory, California Institute of Technology under a contract with NASA). 
This publication made use of information from the Red MSX Source survey database which was constructed with support from the Science and Technology Facilities Council of the UK. 
This paper made use of the Canadian Galactic Plane Survey (supported by a grant from the Natural Sciences and Engineering Research Council of Canada). 
This publication made use of data products from the European Space Agency (ESA) space mission Gaia, the data from which were processed by the Gaia Data Processing and Analysis Consortium (DPAC). The Gaia Archive is reachable from the Gaia home page (\url{http://www.cosmos.esa.int/gaia})

%%%%%%%%%%%%%%%%%%%%%%%%%%%%%%%%%%%%%%%%%%%%%%%%%%
%\iffalse
\section*{Data Availability}

The data supporting the plots within this article are available on reasonable request to the corresponding author.

%\fi


\vspace{1 cm}
\bibliographystyle{mnras}
\bibliography{library} % if your bibtex file is called example.bib

%\begin{thebibliography}{99}
%\end{thebibliography}

%%%%%%%%%%%%%%%%%%%%%%%%%%%%%%%%%%%%%%%%%%%%%%%%%%

%%%%%%%%%%%%%%%%% APPENDICES %%%%%%%%%%%%%%%%%%%%%
\iffalse
\appendix

\section{Some extra material}

If you want to present additional material which would interrupt the flow of the main paper,
it can be placed in an Appendix which appears after the list of references.
\fi
%%%%%%%%%%%%%%%%%%%%%%%%%%%%%%%%%%%%%%%%%%%%%%%%%%

% Don't change these lines
\bsp	% typesetting comment
\label{lastpage}
\end{document}